\pdfoutput=1
\documentclass[final,12pt,nopreprintline]{elsarticle}

\usepackage{amssymb}
\usepackage{amsmath}
\usepackage{lineno}
\usepackage[hidelinks]{hyperref}
\usepackage{cleveref}
\usepackage{enumitem}
\usepackage{caption}
\usepackage{booktabs}
\usepackage{array}
\usepackage{graphicx}
\usepackage{tikz}
\usepackage{float}
\usepackage[section]{placeins}
\usepackage{microtype}

\newcounter{bla}
\newenvironment{refnummer}{%
\list{[\arabic{bla}]}%
{\usecounter{bla}%
 \setlength{\itemindent}{0pt}%
 \setlength{\topsep}{0pt}%
 \setlength{\itemsep}{0pt}%
 \setlength{\labelsep}{2pt}%
 \setlength{\listparindent}{0pt}%
 \settowidth{\labelwidth}{[9]}%
 \setlength{\leftmargin}{\labelwidth}%
 \addtolength{\leftmargin}{\labelsep}%
 \setlength{\rightmargin}{0pt}}}
 {\endlist}

\newcommand{\code}[1]{\texttt{#1}}
\newcommand{\breakableurl}[1]{{\def\UrlBreaks{\do\/\do\-\do\.\do\_\do\a\do\b\do\c\do\d\do\e\do\f\do\0\do\1\do\2\do\3\do\4\do\5\do\6\do\7\do\8\do\9}\url{#1}}}
\newcommand{\longhash}[1]{{\def\UrlBreaks{\do\a\do\b\do\c\do\d\do\e\do\f\do\0\do\1\do\2\do\3\do\4\do\5\do\6\do\7\do\8\do\9}\path{#1}}}
\newcommand{\plotAidDisclosure}{Plotting code was prepared with assistance from
Claude (Anthropic); see \cref{sec:aiWorkflow}.}

\makeatletter
\def\ps@pprintTitle{%
    \let\@oddhead\@empty
    \let\@evenhead\@empty
    \let\@oddfoot\@empty
    \let\@evenfoot\@empty}
\makeatother

\journal{Computer Physics Communications}

\begin{document}

\begin{frontmatter}

\title{\texttt{acousticHelmholtzFoam}: an MPI-parallel unstructured
finite-volume solver for frequency-domain acoustics in heterogeneous two-phase
media with perfectly matched layers}

\author[mma]{Chuanchao Xu}
\ead{xu@mma.tu-darmstadt.de}
\author[mma]{Jun Liu}
\author[mma]{Dieter~Bothe}
\ead{bothe@mma.tu-darmstadt.de}
\author[mma]{Tomislav Maric\texorpdfstring{\corref{cor1}}{}}
\ead{maric@mma.tu-darmstadt.de}
\cortext[cor1]{Corresponding author.}

\affiliation[mma]{organization={Mathematical Modeling and Analysis,
    Mathematics Department, TU Darmstadt},
    addressline={Peter-Gruenberg-Strasse 10},
    city={Darmstadt}, postcode={64287},
    country={Germany}}

\begin{abstract}
\code{acousticHelmholtzFoam} is an OpenFOAM application that computes
time-harmonic acoustic pressure fields in heterogeneous gas--liquid media on
unstructured polyhedral meshes.  The complex pressure is stored as real and
imaginary volume fields, the material coefficients are constructed from a
volume-of-fluid phase fraction with geometric face-fraction averaging, and open
boundaries are truncated by Cartesian perfectly matched layers (PMLs) whose
damping tensor is generated by the companion utility \code{setPMLFields}.  The
two coupled real-valued equations are discretized with the cell-centred
finite-volume operators of OpenFOAM and assembled explicitly, including
processor couplings and boundary contributions, into a distributed PETSc matrix
that is solved by default with the MUMPS sparse direct solver.  The solver
writes pressure amplitude, complex velocity, acoustic radiation pressure and
time-averaged momentum flux for acoustic-force evaluation.  This paper documents
the implemented equations, the case interface, the assembly and solution
algorithm, the build procedure and the test infrastructure.  A layered
gas--liquid--PML case with an analytical solution serves as an end-to-end
regression: on 8000 cells the whole-domain relative complex-pressure error is
$1.197\times10^{-3}$, and serial and MPI solutions on up to eight ranks agree
to within $10^{-12}$ in relative $L_2$ norm.  A single-axis
acoustic-levitator height sweep with constant-volume
spheroidal drops demonstrates the two-phase material input and the force output,
and strong- and problem-size-scaling measurements characterize the
direct-solver workflow.  The numerical formulation and its verification are
presented in a companion paper.

\vspace{1ex}
\noindent\textbf{PROGRAM SUMMARY}

\begin{small}
\raggedright
\noindent
{\em Program Title:} \code{acousticHelmholtzFoam} (with the companion utility
\code{setPMLFields})\\
{\em CPC Library link to program files:} (to be added by Technical Editor)\\
{\em Developer's repository link:}
branch \code{feature/acousticLevitation} of the \code{TwoPhaseFlow}
repository, revision \code{bb09dbfb},
\breakableurl{https://github.com/tmaric/TwoPhaseFlow/tree/bb09dbfb34c1830478feecd8dd445957ac15949c}\\
{\em Licensing provisions:} GPLv3\\
{\em Programming language:} C++ (OpenFOAM); POSIX shell and Python~3 for case
drivers, tests and post-processing\\
{\em Nature of problem (approx. 50-250 words):}\\
Time-harmonic acoustic fields in devices such as resonators, transducers and
acoustic levitators are governed by the Helmholtz equation for the complex
pressure amplitude.  When the medium contains both gas and liquid, density and
sound speed jump by three orders of magnitude across the phase interface, the
interface geometry is typically available only as a volume-fraction field on an
unstructured mesh, and the physical domain is open, so the computational domain
must be truncated without spurious reflections.  The program solves this
heterogeneous, boundary-driven Helmholtz problem on polyhedral finite-volume
meshes, with material coefficients derived from a volume-of-fluid phase
fraction and with perfectly matched layers that absorb outgoing waves.  From the
computed complex pressure it evaluates the acoustic velocity, the acoustic
radiation pressure and the time-averaged momentum flux, which are the inputs for
acoustic radiation-force calculations on drops and particles [1].\\
{\em Solution method (approx. 50-250 words):}\\
The complex pressure is split into real and imaginary parts, which turns the
complex-coordinate PML formulation into a coupled system of two real-valued
equations.  Cell densities and compressibilities are linear volume-fraction
averages, and the face inverse density is formed from a geometric
piecewise-linear interface reconstruction shared by the adjacent cells.  Each
equation is discretized with the cell-centred finite-volume Laplacian and
reaction operators of OpenFOAM [2]; the resulting scalar matrices, their
boundary-condition contributions and the processor-interface couplings of the
decomposed mesh are inserted with global indices into one distributed PETSc
matrix [3] with two unknowns per cell.  The linear system is solved with PETSc
Krylov-subspace infrastructure, by default as a single application of the MUMPS
sparse direct factorization [4]; any PETSc solver and preconditioner can be
selected through PETSc run-time options.  The utility \code{setPMLFields}
evaluates a polynomial damping profile for separable rectangular PMLs and
writes it as a diagonal tensor field from which the solver derives its PML
coefficients.  Derived acoustic fields are computed from the converged pressure
with the standard OpenFOAM gradient operators.\\
{\em Additional comments including restrictions and unusual features (approx.
50-250 words):}\\
The program is an OpenFOAM user application and requires OpenFOAM-v2606
(openfoam.com); OpenFOAM.org releases are not supported.  PETSc with MUMPS and
MPI is built by the supplied script at pinned revisions (PETSc~3.25.0, MUMPS
5.8.2).  The levitator cases require the cfMesh plugin distributed with
OpenFOAM, and post-processing uses Python~3 with NumPy, Matplotlib and VTK.  The
model is linear, inviscid and adiabatic about a quiescent state; the phase
interface and material properties are fixed during a solve.  Only separable
Cartesian rectangular PMLs are implemented (\code{PMLType rectangle}); material
interfaces must lie in the physical domain, and each PML must contain a
homogeneous continuation of the adjacent medium.  The default direct solve is
robust for the indefinite Helmholtz--PML system but its memory and time grow
superlinearly with the number of cells, which limits practical problem sizes to
$\mathcal{O}(10^6)$ unknowns per workstation.  Each non-orthogonal correction
sweep reassembles and refactorizes the matrix.  An experimental interface flux
(\code{plicTransmission}) is selectable but is not part of the validated
configuration.\\

\begin{refnummer}
\item C.~Xu, J.~Liu, J.~H.~D\"orsam, M.~Kupnik, T.~Maric, D.~Bothe, An
unstructured finite-volume Helmholtz method with perfectly matched layers for
heterogeneous two-phase acoustics, arXiv:2610.01593 (2026).
\item H.~G. Weller, G.~Tabor, H.~Jasak, C.~Fureby, A tensorial approach to
computational continuum mechanics using object-oriented techniques, Comput.
Phys. 12 (1998) 620--631.
\item S.~Balay et~al., PETSc/TAO Users Manual, ANL-21/39 Rev.~3.25, Argonne
National Laboratory (2026).
\item P.~R. Amestoy, I.~S. Duff, J.-Y. L'Excellent, J.~Koster, A fully
asynchronous multifrontal solver using distributed dynamic scheduling, SIAM J.
Matrix Anal. Appl. 23 (2001) 15--41.
\end{refnummer}
\end{small}
\end{abstract}

\begin{keyword}
\texttt{acousticHelmholtzFoam} \sep Helmholtz equation \sep unstructured
finite-volume method \sep two-phase acoustics \sep perfectly matched layer
\sep OpenFOAM \sep PETSc
\end{keyword}

\end{frontmatter}

\section{Introduction}
\label{sec:intro}

Frequency-domain formulations are the standard tool when the acoustic response
of a device at a prescribed driving frequency is sought, because they replace
the integration over many acoustic periods by the solution of one Helmholtz
problem.  Their use in gas--liquid configurations is more demanding than in a
single medium: density and sound speed are discontinuous at the phase
interface, the interface geometry is often available only as a volume-fraction
field produced by a volume-of-fluid (VOF) flow simulation on an unstructured
mesh, and the physical domain is open, so the computational domain must be
truncated without generating spurious reflections.  Acoustic levitation of
liquid drops \cite{Andrade2018review,andrade2019numerical} combines all three
requirements: the drop reflects and transmits sound, its shape is described by
a volume fraction, and the levitator radiates into free space.

\code{acousticHelmholtzFoam} is an OpenFOAM \cite{Weller1998} user application
that solves the heterogeneous Helmholtz equation for the complex acoustic
pressure on polyhedral finite-volume meshes.  It stores the complex pressure as
two real volume fields, builds density and compressibility fields from the
liquid volume fraction, evaluates the face material coefficient from a
geometric piecewise-linear interface reconstruction, and represents open
boundaries by Cartesian perfectly matched layers (PMLs).  The coupled
real-valued equations are discretized with the finite-volume operators of
OpenFOAM and assembled, including all processor-interface couplings of a
decomposed mesh, into a distributed PETSc \cite{Balay1997,PETScManual} matrix
that is solved with the MUMPS \cite{Amestoy2001,Amestoy2019} direct solver by
default.  The companion utility \code{setPMLFields} generates the PML damping
tensor from a small set of dictionary entries.  The software is developed on a
branch of the \code{TwoPhaseFlow} framework
\cite{ScheuflerRoenby2023TwoPhaseFlow}, whose geometric VOF infrastructure
\cite{Roenby2016,Scheufler2019} supplies the interface reconstruction.

Finite-volume Helmholtz solvers on unstructured meshes have been reported for
thermoacoustic eigenmodes \cite{Nicoud2007} and coastal-ocean sound propagation
\cite{Long2015}, and complex-coordinate PMLs for the time-harmonic Helmholtz
equation are well established \cite{Harari2000,Bermudez2007,Modave2014PML}.  To
the authors' knowledge, no distributed OpenFOAM application combines a
VOF-derived two-phase material description, a cell-centred Helmholtz--PML
discretization and a PETSc block solve.  The numerical formulation, its
consistency properties and its verification on homogeneous waves, layered
transmission, baffled-piston radiation and rigid-sphere radiation forces are
presented in a companion paper \cite{Xu2026HelmholtzPML}.  The present paper is
the program description: it states the implemented equations only to the
extent needed to identify the solver fields, and then documents the case
interface, the assembly and solution algorithm, the parallel execution model,
the build procedure, the test infrastructure, a reproducible regression case,
an application example and the performance of the direct-solver workflow.

The paper is organized as follows.  \Cref{sec:model} summarizes the equations,
boundary conditions and derived quantities that the software implements.
\Cref{sec:software} describes the program components, the case inputs, the
PML utility, the interface-treatment selection, the assembly into PETSc, the
linear solve, the outputs, parallel execution, installation and testing.
\Cref{sec:verification} presents the layered regression and the serial/MPI
equivalence check, \cref{sec:application} the levitator height sweep, and
\cref{sec:performance} the timing measurements.  \Cref{sec:availability}
states availability, reproducibility and restrictions, and
\cref{sec:conclusions} concludes.

\section{Implemented equations}
\label{sec:model}

This section gives the minimum mathematical context needed to explain the
solver fields and dictionaries.  Derivations, the interface-coefficient
discussion and the unstructured finite-volume consistency analysis are given
in \cite{Xu2026HelmholtzPML}.

\subsection{Heterogeneous Helmholtz equation}
\label{sec:helmholtz}

For a quiescent inviscid medium with spatially varying density $\rho$ and
sound speed $c$, the first-order pressure perturbation $p_1$ satisfies the
heterogeneous wave equation
\begin{equation}
    \frac{\partial^2 p_1}{\partial t^2}
    =
    \rho c^2\nabla\cdot\left(\frac{1}{\rho}\nabla p_1\right).
    \label{eq:waveEquation}
\end{equation}
The solver uses the time-harmonic ansatz
\begin{equation}
    p_1(\mathbf{x},t)=\mathrm{Re}\left(e^{-i\omega t}P(\mathbf{x})\right),
    \qquad
    P=P_{re}+iP_{im},
    \label{eq:harmonicAnsatz}
\end{equation}
with angular frequency $\omega=2\pi f$.  Substitution into
\cref{eq:waveEquation} gives the heterogeneous Helmholtz equation
\begin{equation}
    \rho\nabla\cdot\left(\frac{1}{\rho}\nabla P\right)
    + k_\alpha^2 P = 0,
    \qquad
    k_\alpha^2 = \omega^2\rho\beta,
    \qquad
    \beta=\frac{1}{\rho c^2},
    \label{eq:helmholtz}
\end{equation}
where $\beta$ is the local acoustic compressibility, so that $\rho\beta=1/c^2$.
The complex velocity amplitude follows from the first-order momentum equation,
$\mathbf{V}=\nabla P/(i\omega\rho)$, with real and imaginary parts
\begin{equation}
    \mathbf{V}_{re}=\frac{\nabla P_{im}}{\omega\rho},
    \qquad
    \mathbf{V}_{im}=-\frac{\nabla P_{re}}{\omega\rho}.
    \label{eq:complexVelocity}
\end{equation}

\subsection{Two-phase material coefficients}
\label{sec:material}

The material indicator is the liquid volume fraction $\alpha$, with $\alpha=1$
in the liquid, $\alpha=0$ in the gas and intermediate values in interface
cells.  The cell density and compressibility are the linear averages
\begin{equation}
    \rho = \alpha\rho_l + (1-\alpha)\rho_g,
    \qquad
    \beta = \alpha\beta_l + (1-\alpha)\beta_g,
    \qquad
    \beta_i = \frac{1}{\rho_i c_i^2},\quad i\in\{l,g\}.
    \label{eq:materialAverages}
\end{equation}
The pressure-gradient flux across a face $f$ is scaled by the face inverse
density
\begin{equation}
    a_f = \frac{1}{\alpha_f\rho_l+(1-\alpha_f)\rho_g},
    \label{eq:faceInverseDensity}
\end{equation}
where $\alpha_f$ is the liquid-covered area fraction of the face.  In the
validated configuration, $\alpha_f$ is obtained by cutting the face with the
piecewise-linear interface reconstructions of both adjacent cells and averaging
the two candidate fractions; a single valid reconstruction is used directly,
and faces between pure cells take the value zero, one, or the linear
interpolation of $\alpha$ when the two cells belong to different phases.  This
face fraction does not depend on the direction of any velocity field.  Its
definition, its processor-boundary treatment and its effect on reproducibility
are documented in \cite{Xu2026HelmholtzPML}; \cref{sec:interfaceSelection}
explains how it is selected in a case.

\subsection{Perfectly matched layer}
\label{sec:pml}

Open boundaries are represented by adding an absorbing layer to the physical
domain.  The PML is obtained by complex coordinate stretching with a
directional damping tensor $\boldsymbol{\sigma}=\mathrm{diag}(\sigma_x,\sigma_y,\sigma_z)$.
In the physical domain $\boldsymbol{\sigma}=0$.  Inside the PML the
components are non-positive: with the $e^{-i\omega t}$ convention of
\cref{eq:harmonicAnsatz}, $\sigma_j\le 0$ produces attenuation of outgoing
waves.  Throughout this paper the peak damping is quoted as a magnitude
$\sigma_{\max}>0$; the sign is applied by the software
(\cref{sec:setPMLFields}).

The software constructs four coefficient fields from $\boldsymbol{\sigma}$ and
the local material coefficients: two scalar reaction coefficients,
\begin{equation}
\begin{aligned}
    C_0 &= \rho\beta\left(\sigma_x\sigma_y+\sigma_x\sigma_z+\sigma_y\sigma_z\right),\\
    C_1 &= \rho\beta\left[
        \omega(\sigma_x+\sigma_y+\sigma_z)
        -\frac{\sigma_x\sigma_y\sigma_z}{\omega}
    \right],
\end{aligned}
\label{eq:pmlReactionCoefficients}
\end{equation}
and two diagonal tensors $\mathbf{T}_0=\mathrm{diag}(T_{0xx},T_{0yy},T_{0zz})$,
$\mathbf{T}_1=\mathrm{diag}(T_{1xx},T_{1yy},T_{1zz})$ whose entries, for each
Cartesian direction $i$ with $j$ and $l$ denoting the remaining two directions,
are
\begin{equation}
\begin{aligned}
    T_{0ii} &=
    \frac{\sigma_i(\sigma_j+\sigma_l-\sigma_i)-\sigma_j\sigma_l}
    {\omega^2+\sigma_i^2},\\
    T_{1ii} &=
    \frac{(\sigma_x\sigma_y\sigma_z)/\omega
    +\omega(\sigma_j+\sigma_l-\sigma_i)}
    {\omega^2+\sigma_i^2}.
\end{aligned}
\label{eq:pmlTensorCoefficients}
\end{equation}
Since $\rho\beta=1/c^2$, \cref{eq:pmlReactionCoefficients} coincides with the
$1/c^2$-scaled coefficients of \cite{Xu2026HelmholtzPML}; the software
evaluates them cell by cell from the volume-fraction-averaged $\rho$ and
$\beta$.  With these fields, the real-valued PML form of the Helmholtz equation
is
\begin{equation}
\begin{aligned}
    \hat A P_{re} + \hat B P_{im} &= 0, \\
    \hat A P_{im} - \hat B P_{re} &= 0,
\end{aligned}
\label{eq:pmlSplit}
\end{equation}
with the diagonal operator
\begin{equation}
    \hat A =
    \rho\nabla\cdot\left(\frac{1}{\rho}\nabla\right)
    + \nabla\cdot(\mathbf{T}_0\cdot\nabla)
    + (k_\alpha^2-C_0)
    \label{eq:diagonalOperator}
\end{equation}
and the real--imaginary coupling operator
\begin{equation}
    \hat B = \nabla\cdot(\mathbf{T}_1\cdot\nabla)+C_1.
    \label{eq:couplingOperator}
\end{equation}
The tensor fluxes are written without the $\rho(\cdot)/\rho$ scaling because
material interfaces are confined to the physical domain, where
$\mathbf{T}_0=\mathbf{T}_1=0$, and each PML contains a homogeneous
continuation of the adjacent medium, so that $\rho\beta=1/c_{\mathrm{ext}}^2$
inside the layer.  These are modelling restrictions of the software, not
merely of the supplied cases.

\subsection{Boundary conditions}
\label{sec:boundaryConditions}

\Cref{eq:pmlSplit} is homogeneous; the acoustic field is driven entirely by the
boundary conditions, which therefore form the right-hand side of the discrete
system (\cref{sec:blockSystem}).  A prescribed complex normal velocity
$V_n=V_{n,re}+iV_{n,im}$, defined with the outward normal of the fluid domain,
enters through the momentum equation as the pair of Neumann conditions
\begin{equation}
    \frac{\partial P_{re}}{\partial n}=-\omega\rho V_{n,im},
    \qquad
    \frac{\partial P_{im}}{\partial n}=\omega\rho V_{n,re}.
    \label{eq:velocityBC}
\end{equation}
A transducer face moving into the fluid with real velocity amplitude $u_0$ has
$V_{n,re}=-u_0$ and is imposed as $\partial P_{im}/\partial n=-\omega\rho u_0$
with $\partial P_{re}/\partial n=0$.  In the supplied cases this is written as
an OpenFOAM \code{fixedGradient} condition on \code{Pim}, whose gradient is
evaluated from the dictionary entries for $f$, $\rho$ and $u_0$
(\cref{sec:caseInputs}), and a \code{zeroGradient} condition on \code{Pre}.
Sound-hard walls and the outer boundary of every PML use \code{zeroGradient}
for both components; the PML outer condition $\partial_n P=0$ is the truncation
used in the analytical reference of the regression case.  Dirichlet data can
be imposed with \code{fixedValue} on either component.  Axisymmetric wedge
cases use \code{wedge} and \code{symmetryPlane} patches, and one- and
two-dimensional cases use \code{empty} patches, all of which are handled by
the standard OpenFOAM patch-field machinery.

\subsection{Discrete block system}
\label{sec:blockSystem}

The solver discretizes \cref{eq:pmlSplit} with the cell-centred unstructured
finite-volume operators of OpenFOAM \cite{Jasak1996,Moukalled2016}.  For either
pressure component $Q$, the diagonal block is assembled from
\begin{equation}
\begin{aligned}
    &\texttt{rho*fvm::laplacian(invRhof, Q)}
    +\texttt{fvm::laplacian(T0, Q)}\\
    &\quad+\texttt{fvm::Sp(k2 - C0, Q)},
\end{aligned}
    \label{eq:diagonalAssembly}
\end{equation}
and the coupling block from the two separate scalar matrices
\begin{equation}
    \texttt{fvm::laplacian(T1, Q)}
    \qquad\text{and}\qquad
    \texttt{fvm::Sp(C1, Q)},
    \label{eq:couplingAssembly}
\end{equation}
where \code{invRhof} is the face field of \cref{eq:faceInverseDensity} and
\code{k2} is $k_\alpha^2$.  The orthogonal part of each face gradient and the
reaction terms are implicit; the non-orthogonal correction of the selected
Laplacian scheme is explicit and is refreshed in the correction loop of
\cref{sec:linearSolve}.  After assembly over all cells the discrete equations
form the real-valued block system
\begin{equation}
\begin{bmatrix}
    A_h & B_h \\
    -B_h & A_h
\end{bmatrix}
\begin{bmatrix}
    \mathbf{P}_{re} \\ \mathbf{P}_{im}
\end{bmatrix}
=
\begin{bmatrix}
    \mathbf{b}_{re} \\ \mathbf{b}_{im}
\end{bmatrix},
\label{eq:blockSystem}
\end{equation}
in which $A_h$ and $B_h$ are the matrices of \cref{eq:diagonalAssembly} and
\cref{eq:couplingAssembly}, and the right-hand side collects the source and
boundary-condition contributions of each scalar matrix, combined with the same
signs as the corresponding blocks.  With the Neumann data of
\cref{eq:velocityBC}, $\mathbf{b}_{im}$ carries the transducer forcing.  The
solver stores the unknowns in interleaved order, cell $g$ occupying rows $2g$
(\code{Pim}) and $2g+1$ (\code{Pre}) of the global PETSc vector; this ordering
keeps the unknowns of each rank contiguous and is equivalent to
\cref{eq:blockSystem} up to a permutation.

\subsection{Derived acoustic quantities}
\label{sec:derivedQuantities}

After each solve the software evaluates, with the selected OpenFOAM gradient
scheme, the velocity components of \cref{eq:complexVelocity}, the pressure
amplitude $p_a=\sqrt{P_{re}^2+P_{im}^2}$, the time-averaged second-order
pressure
\begin{equation}
    p_{\mathrm{rad}}
    =
    \frac{\beta}{4}\left(P_{re}^2+P_{im}^2\right)
    -
    \frac{\rho}{4}\left(\mathbf{V}_{re}\cdot\mathbf{V}_{re}
    +\mathbf{V}_{im}\cdot\mathbf{V}_{im}\right),
    \label{eq:radiationPressure}
\end{equation}
and the time-averaged acoustic momentum-flux tensor
\begin{equation}
    \boldsymbol{\Pi}
    =\langle\rho\,\mathbf{v}_1\mathbf{v}_1\rangle
    =\frac{\rho}{2}\left(\mathbf{V}_{re}\mathbf{V}_{re}
    +\mathbf{V}_{im}\mathbf{V}_{im}\right).
    \label{eq:momentumFlux}
\end{equation}
The acoustic radiation force on a closed surface $\partial\Omega_p$ with
normal $\mathbf{n}$ pointing out of the fluid is
\begin{equation}
    \mathbf{F}
    =\int_{\partial\Omega_p}
    \left(p_{\mathrm{rad}}\mathbf{I}+\boldsymbol{\Pi}\right)\cdot\mathbf{n}\,\mathrm{d}A ,
    \label{eq:radiationForce}
\end{equation}
which is the integral evaluated in \cite{Xu2026HelmholtzPML} for the
rigid-sphere benchmark.  On a sound-hard wall the normal component of
$\boldsymbol{\Pi}\cdot\mathbf{n}$ vanishes, so the normal wall force reduces
to the integral of $p_{\mathrm{rad}}$; this is the form used by the
reflector-force function object of the application case
(\cref{sec:application}).  The expression \cref{eq:radiationPressure} is
derived for a homogeneous medium with constant $\rho$ and $c$
\cite{Xu2026HelmholtzPML,bruus2012acoustofluidicsARF}.  The software
evaluates it with the local volume-fraction-averaged $\rho$ and $\beta$ of
\cref{eq:materialAverages}, so that its values inside interface cells of a
two-phase case are an approximation and should be used only where the medium
is locally homogeneous, for instance on rigid walls in the gas.

\section{Software description}
\label{sec:software}

\subsection{Components and repository layout}
\label{sec:components}

\begin{figure}[htbp]
    \centering
    \begin{tikzpicture}[x=1cm,y=1cm,>=latex,font=\footnotesize]
        \tikzset{flow/.style={draw,align=center,minimum height=0.95cm,
            text width=2.55cm,inner sep=4pt}}
        \node[flow] (input) at (0,0)
            {mesh, \code{alpha.water},\\dictionaries, BCs};
        \node[flow] (pml) at (3.45,0)
            {\code{setPMLFields}\\writes \code{sigma}};
        \node[flow] (assembly) at (6.90,0)
            {material and PML\\coefficients, FVM\\matrices};
        \node[flow] (solve) at (10.35,0)
            {PETSc/MUMPS\\distributed solve};
        \node[flow] (output) at (6.90,-1.75)
            {\code{Pre}, \code{Pim}, \code{pa},\\
             \code{Ure}, \code{Uim}, \code{pr}, \code{momFlux}};
        \node[flow] (post) at (3.45,-1.75)
            {function objects,\\case scripts};

        \draw[->,thick] (input) -- (pml);
        \draw[->,thick] (pml) -- (assembly);
        \draw[->,thick] (assembly) -- (solve);
        \draw[->,thick] (solve.south) |- (output.east);
        \draw[->,thick] (output) -- (post);
    \end{tikzpicture}
    \caption{Case workflow from OpenFOAM case data through PML generation,
    finite-volume assembly, distributed solution and written acoustic fields.
    \code{setPMLFields} and \code{acousticHelmholtzFoam} are the two
    executables of the program; the remaining steps use standard OpenFOAM
    utilities.}
    \label{fig:workflow}
\end{figure}
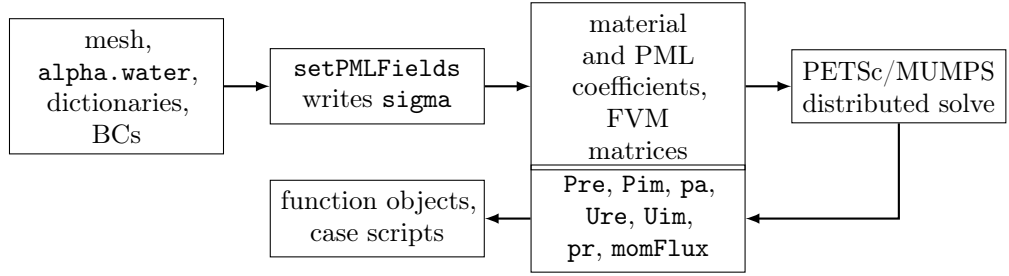

The program is developed on the \code{feature/acousticLevitation} branch of
the \code{TwoPhaseFlow} repository \cite{ScheuflerRoenby2023TwoPhaseFlow} and
reuses its geometric VOF libraries.  \Cref{tab:components} lists the
components that form the program and the case material discussed in this
paper.  The complete case workflow is summarized in \cref{fig:workflow}.

\begin{table}[htbp]
    \centering
    \caption{Program components in the \code{TwoPhaseFlow} repository
    (revision \code{bb09dbfb}).}
    \label{tab:components}
    \small
    \begin{tabular}{@{}>{\raggedright\arraybackslash}p{0.44\textwidth}>{\raggedright\arraybackslash}p{0.52\textwidth}@{}}
        \toprule
        Path & Content \\
        \midrule
        \path{solver/acousticHelmholtzFoam/} &
        Solver sources: \code{acousticHelmholtzFoam.C}, \code{createFields.H},
        \code{readTransportProperties.H}, \code{computePMLCoefs.H},
        \code{petscBlockAssembly.H}, \code{petscBlockSolve.H}, \code{Make/}.\\
        \path{apps/utilities/mesh/setPMLFields/} &
        PML damping utility (driver); the profile and dictionary parsing are
        in \path{src/acousticPML/PMLFieldSetup.H}.\\
        \path{src/acousticInterface/} &
        Library selecting the face-fraction and flux treatment
        (\cref{sec:interfaceSelection}); \code{legacy/} preserves the original
        donor-based face-fraction code; \code{README.md} documents the options.\\
        \path{apps/utilities/preProcessing/setAlphaField/} &
        Repository version of \code{setAlphaField}, initializing
        \code{alpha.water} from an implicit surface (plane, sphere, ellipsoid).\\
        \path{apps/benchmark/testAcousticInterface/} &
        Compiled unit checks of the geometric face-fraction and flux code.\\
        \path{run/acousticTests/FrequencyDomainTests/layeredInterface1D/} &
        Regression case of \cref{sec:layeredRegression} with \code{Allrun},
        \code{caseParams.sh}, \code{prepareCase} and post-processing scripts.\\
        \path{run/acousticTests/FrequencyDomainTests/Andrade2019/WedgeLevitatorDropAlpha/} &
        Application case of \cref{sec:application} with \code{runHeightSweep.sh}
        and \code{runPerformanceScaling.sh}.\\
        \path{testsuite/acousticInterface/} &
        \code{pytest} module, comparison drivers and verification records.\\
        \path{petsc_build.sh}, \path{scripts/bashrc},
        \path{scripts/check-acoustic-environment.sh} &
        Dependency build, environment setup and environment check.\\
        \path{INSTALL_ACOUSTICS.md}, \path{SUBMISSION_ACOUSTICS.md} &
        Installation guide and configuration record of the published cases.\\
        \bottomrule
    \end{tabular}
\end{table}

\subsection{Case structure and required inputs}
\label{sec:caseInputs}

A case follows the usual OpenFOAM layout.  \Cref{tab:caseInputs} lists the
files the solver reads.  The pressure components \code{Pre} and \code{Pim}
carry the boundary conditions of \cref{sec:boundaryConditions}; the damping
tensor \code{sigma} is written by \code{setPMLFields} and must exist in the
start-time directory; the volume fraction \code{alpha.water} is optional and
defaults to zero, which makes the whole domain gas.  Several further fields
(\code{pa}, \code{U}, \code{Ure}, \code{Uim}, \code{alphaf}, \code{rho}) are
declared with the read flag \code{MUST\_READ} in \code{createFields.H} but are
constructed from default values or expressions, so the flag is not enforced
and the fields need not be supplied; at start-up the solver prints one
OpenFOAM warning per such field (``The readOption \code{MUST\_READ} or
\code{READ\_MODIFIED} suggests that a read constructor for field \ldots{}
would be more appropriate''), which can be ignored.  The zero velocity field
\code{U} exists only to instantiate the \code{isoAdvector} object that
provides the interface reconstruction.

\begin{table}[htbp]
    \centering
    \caption{Case inputs read by \code{acousticHelmholtzFoam}.}
    \label{tab:caseInputs}
    \small
    \begin{tabular}{@{}>{\raggedright\arraybackslash}p{0.30\textwidth}>{\raggedright\arraybackslash}p{0.13\textwidth}>{\raggedright\arraybackslash}p{0.52\textwidth}@{}}
        \toprule
        File & Required & Content \\
        \midrule
        \code{0/Pre}, \code{0/Pim} & yes & Real and imaginary pressure with
        boundary conditions (\cref{sec:boundaryConditions}); the internal
        field is the initial guess and is normally zero.\\
        \code{0/sigma} & yes & Diagonal damping tensor $[\mathrm{s^{-1}}]$
        written by \code{setPMLFields}.\\
        \code{0/alpha.water} & no & Liquid volume fraction; absent means
        $\alpha=0$ everywhere.  Written by \code{setAlphaField} or taken from
        a VOF simulation.\\
        \path{constant/transportProperties} & yes & Material, frequency and
        PML parameters (\cref{tab:transportProperties}).\\
        \path{constant/polyMesh} & yes & Mesh; processor directories after
        \code{decomposePar}.\\
        \path{system/controlDict} & yes & \code{application acousticHelmholtzFoam};
        \code{endTime}/\code{deltaT} set the number of outer iterations
        (\cref{sec:linearSolve}); \code{functions} may evaluate forces.\\
        \path{system/fvSchemes} & yes & \code{gradSchemes},
        \code{laplacianSchemes}, \code{snGradSchemes} and the optional
        \code{acousticInterface} sub-dictionary (\cref{sec:interfaceSelection}).\\
        \path{system/fvSolution} & yes & \code{solvers."alpha.*"} with the
        reconstruction scheme (\code{plicRDF}) and tolerances used by the
        interface reconstruction; \code{SIMPLE.nNonOrthogonalCorrectors}.\\
        \path{system/decomposeParDict} & MPI only & Decomposition method and
        rank count.\\
        \bottomrule
    \end{tabular}
\end{table}

\Cref{tab:transportProperties} lists the entries of
\code{constant/transportProperties}.  The compressibilities \code{kg} and
\code{kl} are read as independent entries; the supplied cases compute them
from the densities and sound speeds with OpenFOAM's \code{\#eval} syntax, so
that only $\rho_g$, $\rho_l$, $c_g$, $c_l$ and $f$ are edited.  The sound
speeds \code{cg} and \code{cl} are read but are not used by the solver
itself.

\begin{table}[htbp]
    \centering
    \caption{Entries of \code{constant/transportProperties} read by the
    program.  Frequency, densities and compressibilities are read by the
    solver; the PML entries are read by \code{setPMLFields}.}
    \label{tab:transportProperties}
    \small
    \begin{tabular}{@{}llp{0.60\textwidth}@{}}
        \toprule
        Entry & Unit & Meaning \\
        \midrule
        \code{f} & Hz & Driving frequency.\\
        \code{rhog}, \code{rhol} & kg\,m$^{-3}$ & Gas and liquid density.\\
        \code{kg}, \code{kl} & Pa$^{-1}$ & Gas and liquid compressibility,
        $\beta_i=1/(\rho_ic_i^2)$.\\
        \code{cg}, \code{cl} & m\,s$^{-1}$ & Gas and liquid sound speed (read,
        not used by the solver).\\
        \code{PMLType} & -- & Must be \code{rectangle}; any other value is a
        fatal error.\\
        \code{L} & m & PML thickness, read as a plain scalar.\\
        \code{po} & -- & Polynomial order of the damping profile; the legacy
        key \code{p} is accepted; default 2.\\
        \code{sigmaMax} & s$^{-1}$ & Peak damping magnitude, $\ge0$; negated
        internally.  Default 0.\\
        \code{sigmaMaxVector} & s$^{-1}$ & Optional signed per-direction peak
        damping; overrides \code{sigmaMax}.  A zero component disables damping
        in that direction.\\
        \code{PMLMax}, \code{PMLMin} & m & Outer bounds of the rectangular
        PML box in each direction.\\
        \bottomrule
    \end{tabular}
\end{table}

\subsection{PML generation with \code{setPMLFields}}
\label{sec:setPMLFields}

\code{setPMLFields} reads \code{constant/transportProperties}, evaluates the
damping tensor at every cell centre and physical boundary-face centre, and
writes the diagonal tensor field \code{sigma} into the start-time directory.
For each Cartesian component $i$ the distances $d_i^{+}=x_i^{\max}-x_i$ and
$d_i^{-}=x_i-x_i^{\min}$ to the outer bounds \code{PMLMax} and \code{PMLMin}
are evaluated.  If either distance is smaller than the thickness $L$, the
damping in that direction is
\begin{equation}
    \sigma_i(\mathbf{x})
    =
    \bar{\sigma}_i
    \left[
    \min\left(\max\left(\frac{L-d_i}{L},0\right),1\right)
    \right]^{p_o},
    \qquad
    d_i=\min(d_i^{+},d_i^{-}),
    \label{eq:dampingProfile}
\end{equation}
and zero otherwise, where $\bar{\sigma}_i\le0$ is the $i$th component of
\code{sigmaMaxVector} and defaults to $-\code{sigmaMax}$ in all directions.
This produces the separable tensor
$\boldsymbol{\sigma}=\mathrm{diag}(\sigma_x(x),\sigma_y(y),\sigma_z(z))$,
zero in the physical domain and increasing smoothly to the prescribed peak at
the outer boundary.  Corner regions where two or three components are
non-zero are handled by the coupling terms of \cref{eq:pmlReactionCoefficients}
and \cref{eq:pmlTensorCoefficients}.  Setting one component of
\code{sigmaMaxVector} to zero disables damping in that direction, which the
axisymmetric wedge cases use for the wedge direction.  The damping is
coordinate-wise; a radial or spherical PML is not available.  \code{empty},
\code{wedge} and \code{processor} patches are left at zero and are otherwise
handled by their patch types.  The damping profile is not tuned automatically;
the influence of $\sigma_{\max}$, $L$ and mesh resolution on the truncation
error is quantified in \cite{Xu2026HelmholtzPML}, and \cref{sec:layeredRegression}
states the values used for the regression case.

At solver start-up, \code{sigma} is read and the fields \code{C0}, \code{C1},
\code{T0} and \code{T1} are computed from \cref{eq:pmlReactionCoefficients}
and \cref{eq:pmlTensorCoefficients} for all cells and boundary faces, using
the local $\rho\beta$; the denominators $\omega^2+\sigma_i^2$ are guarded
against division by zero.  These four fields are written with the solution
for inspection.

\subsection{Interface treatment selection}
\label{sec:interfaceSelection}

The face fraction $\alpha_f$ and the face flux are selected at start-up from
the optional \code{acousticInterface} sub-dictionary of \code{system/fvSchemes}.
The published cases specify
\begin{verbatim}
acousticInterface
{
    areaFraction       plicAverage;
    flux               legacy;
    writeDiagnostics   false;
}
\end{verbatim}
\Cref{tab:interfaceOptions} lists the accepted combinations.  Both entries
default to \code{legacy} when the sub-dictionary or an entry is missing, so
cases prepared for earlier revisions retain the original behaviour; the
published cases therefore set both entries explicitly.  Unknown method names
and the combination \code{legacy}/\code{plicTransmission} are rejected with a
fatal error.  With \code{areaFraction plicAverage} the geometric face
fractions are computed from the planar reconstructions of the adjacent cells
as described in \cref{sec:material}; on processor faces the plane
descriptions and their availability are exchanged so that both ranks evaluate
the same fraction, and the agreement is checked to an absolute tolerance of
$10^{-12}$.  A mixed cell without a valid reconstruction, non-finite geometry
and unsupported patch types (moving meshes, cyclic, AMI and overset
configurations) are errors.  With \code{writeDiagnostics true}, per-rank
tab-separated files of face fractions, oriented phase areas and reconstructed
planes are written under \code{postProcessing/acousticInterface/}.

\begin{table}[htbp]
    \centering
    \caption{Accepted \code{acousticInterface} settings.}
    \label{tab:interfaceOptions}
    \small
    \begin{tabular}{@{}llp{0.52\textwidth}@{}}
        \toprule
        \code{areaFraction} & \code{flux} & Meaning \\
        \midrule
        \code{legacy} & \code{legacy} & Original donor-based PLIC face areas,
        selected from the sign of a face flux, with the one-field flux of
        \cref{eq:faceInverseDensity} (default when entries are missing).\\
        \code{plicAverage} & \code{legacy} & Geometric face fractions averaged
        over the available adjacent reconstructions with the same one-field
        flux.  \textbf{Validated configuration} used for all published
        results.\\
        \code{plicAverage} & \code{plicTransmission} & Geometric face fractions
        with an experimental multipoint transmission flux; opt-in, not
        validated, controlled by the sub-dictionary
        \code{plicTransmissionCoeffs}.\\
        \bottomrule
    \end{tabular}
\end{table}

The reconstruction itself is provided by the \code{isoAdvector} and
\code{plicRDF} classes \cite{Roenby2016,Scheufler2019} through the
\code{solvers."alpha.*"} entry of \code{system/fvSolution}; the tolerances
\code{surfCellTol} and \code{isoFaceTol} classify pure and mixed cells.

\subsection{Assembly of the distributed system}
\label{sec:assembly}

The solver assembles \cref{eq:blockSystem} explicitly into a PETSc matrix of
type \code{MATAIJ} (\code{MPIAIJ} in parallel) rather than delegating the
solve to the OpenFOAM linear-solver framework, because the real--imaginary
coupling blocks and the PML tensor operators have no counterpart in the
segregated OpenFOAM solvers.  The steps are:
\begin{enumerate}[leftmargin=*]
    \item Global cell numbering.  A \code{globalIndex} object assigns each
    cell a global index; rank $r$ owns a contiguous range, and the two rows of
    cell $g$ are $2g$ and $2g+1$.  The global indices of the cells on the far
    side of every processor patch are exchanged once, since they depend only
    on the mesh topology; the exchange uses non-blocking transfers so that
    large processor patches cannot deadlock.
    \item Preallocation.  Each row receives two diagonal-block entries per
    cell and two per internal face for owner and neighbour; faces on processor
    patches contribute to the off-process count.  Additional columns required
    by the experimental transmission operators are counted as a superset.  PETSc
    is instructed to treat any insertion outside the preallocated pattern as
    an error, so an assembly defect fails loudly.
    \item Scalar operators.  For each component the \code{fvScalarMatrix}
    objects of \cref{eq:diagonalAssembly} and \cref{eq:couplingAssembly} are
    built with the schemes of \code{system/fvSchemes}.  Their diagonal,
    upper and lower coefficients are inserted with global row and column
    indices; the coupling matrices enter the off-diagonal blocks with the
    signs of \cref{eq:blockSystem}.  The two coupling matrices are kept
    separate because the Laplacian contributes face couplings whereas the
    reaction term is purely diagonal.
    \item Boundary conditions.  Physical boundary conditions contribute an
    internal coefficient and a source.  OpenFOAM adds the former to the
    matrix diagonal only inside its own \code{solve()} call, so the solver
    adds it explicitly; the source enters the right-hand side.
    \item Processor coupling.  The coefficients that OpenFOAM would apply
    implicitly through its \code{processorLduInterface} objects are inserted
    as explicit off-process matrix entries using the exchanged neighbour
    indices, so that PETSc sees the identical operator on the decomposed mesh
    as on a single rank.
    \item Right-hand side.  $\mathbf{b}_{im}$ and $\mathbf{b}_{re}$ are formed
    from the sources of the scalar matrices, including their boundary
    contributions, with the signs of \cref{eq:blockSystem}, and scattered
    into the PETSc vector in the interleaved ordering.
    \item Transmission operators.  When the experimental transmission flux is
    selected, the one-field coefficient of \cref{eq:faceInverseDensity} is
    masked on the replaced faces and the face operators are inserted into
    both diagonal blocks.  With the validated \code{legacy} flux this list is
    empty and the step is a no-op.
\end{enumerate}
Because the assembled operator is identical on any number of ranks up to
floating-point summation order, serial and parallel solutions agree to
round-off; this is the property tested in \cref{sec:parallelEquivalence}.

\subsection{Linear solve and iteration control}
\label{sec:linearSolve}

The assembled system is solved with a PETSc \code{KSP} object.  If the user
supplies no PETSc options, the solver sets \code{-ksp\_type preonly},
\code{-pc\_type lu} and \code{-pc\_factor\_mat\_solver\_type mumps}, that is,
a single application of the MUMPS parallel sparse direct factorization
\cite{Amestoy2001,Amestoy2019}.  All PETSc options are honoured through
\code{KSPSetFromOptions}, so an iterative Krylov method and preconditioner
from the PETSc build (the supplied build includes Hypre, SuperLU\_DIST and
ScaLAPACK in addition to MUMPS) can be selected without recompilation.  PETSc
options are read from the standard PETSc option sources, in particular the
environment variable \code{PETSC\_OPTIONS} and a \code{.petscrc} file, because
the OpenFOAM argument parser owns the command line: an invocation such as
\code{acousticHelmholtzFoam -ksp\_type gmres} is rejected by OpenFOAM with
``Invalid option'', whereas the same options exported in the environment,
{\small
\begin{verbatim}
PETSC_OPTIONS="-ksp_type gmres -pc_type hypre" acousticHelmholtzFoam
\end{verbatim}}
\noindent are honoured.  In that particular test on the regression case of
\cref{sec:layeredRegression}, GMRES preconditioned by BoomerAMG broke down
after 60 iterations and the solver stopped with the fatal diagnostic
described below, which illustrates both the difficulty of the indefinite
Helmholtz--PML system for standard algebraic preconditioners and the
solver's refusal to write an unconverged field.

After every solve the program checks the \code{KSP} convergence reason and
aborts on failure, computes the true residual $\mathbf{r}=\mathbf{M}\mathbf{x}-\mathbf{b}$
of the assembled system, rejects non-finite results, and prints the relative
residual $\|\mathbf{r}\|_2/\|\mathbf{b}\|_2$ and the normwise backward error
$\|\mathbf{r}\|_\infty/(\|\mathbf{M}\|_\infty\|\mathbf{x}\|_\infty+\|\mathbf{b}\|_\infty)$.
The solution is scattered back into \code{Pim} and \code{Pre} and their
boundary conditions are updated.

The outer structure reuses OpenFOAM's steady-state \code{simpleControl}.
Each pseudo-time step of \code{controlDict} (\code{endTime}/\code{deltaT}
steps; the supplied cases use one) performs
\code{nNonOrthogonalCorrectors}+1 correction sweeps.  In each sweep the matrix
and right-hand side are zeroed and reassembled with the current explicit
non-orthogonal correction, the system is solved, and the relative change
$\|\Delta\mathbf{P}\|_2/\|\mathbf{P}\|_2$ of the combined pressure vector with
respect to the previous sweep is printed.  On orthogonal meshes, as in the
regression case, \code{nNonOrthogonalCorrectors 0} yields exactly one
assembly and one direct solve.  On non-orthogonal meshes each additional sweep
costs a complete refactorization with the default direct solver.  The
pseudo-time is not physical time; the driving frequency enters only through
\code{f}.

\subsection{Output fields}
\label{sec:outputs}

At each write time the solver writes the fields of \cref{tab:outputs} in the
selected \code{writeFormat}.  The derived quantities are cell-centred and use
the gradient scheme of \code{fvSchemes}; their accuracy on distorted meshes is
discussed in \cite{Xu2026HelmholtzPML}.  Surface integrals of \code{pr} and
\code{momFlux} over patches, as needed for \cref{eq:radiationForce}, are left
to function objects or post-processing so that the force definition remains
under user control.

\begin{table}[htbp]
    \centering
    \caption{Fields written by \code{acousticHelmholtzFoam}.}
    \label{tab:outputs}
    \small
    \begin{tabular}{@{}llp{0.58\textwidth}@{}}
        \toprule
        Field & Type & Definition \\
        \midrule
        \code{Pre}, \code{Pim} & scalar & Real and imaginary pressure
        $[\mathrm{Pa}]$.\\
        \code{pa} & scalar & Pressure amplitude $\sqrt{P_{re}^2+P_{im}^2}$.\\
        \code{Ure}, \code{Uim} & vector & Velocity components of
        \cref{eq:complexVelocity} $[\mathrm{m\,s^{-1}}]$.\\
        \code{pr} & scalar & Radiation pressure $p_{\mathrm{rad}}$ of
        \cref{eq:radiationPressure} $[\mathrm{Pa}]$.\\
        \code{momFlux} & tensor & Momentum-flux tensor $\boldsymbol{\Pi}$ of
        \cref{eq:momentumFlux} $[\mathrm{Pa}]$.\\
        \code{C0}, \code{C1}, \code{T0}, \code{T1} & scalar, tensor & PML
        coefficients of \cref{eq:pmlReactionCoefficients} and
        \cref{eq:pmlTensorCoefficients}.\\
        \code{sigma} & tensor & Damping tensor (rewritten unchanged).\\
        \code{U} & vector & Zero velocity used to instantiate the interface
        reconstruction.\\
        \bottomrule
    \end{tabular}
\end{table}

\subsection{Parallel execution}
\label{sec:parallel}

Parallel runs follow the standard OpenFOAM pattern: \code{decomposePar}
partitions the mesh (the supplied cases use Scotch \cite{Chevalier2008}),
\code{mpirun -np N acousticHelmholtzFoam -parallel} runs the solver, and
\code{reconstructPar} merges the fields.  \code{setPMLFields} is executed on
the undecomposed mesh before decomposition because \code{sigma} is a
decomposed input like any other field.  Inter-rank communication occurs in
three places: the exchange of reconstructed-plane descriptions for the face
fractions on processor faces (\cref{sec:interfaceSelection}), the one-time
exchange of neighbour global indices and the insertion of off-process matrix
entries (\cref{sec:assembly}), and the internal communication of PETSc and
MUMPS during factorization and solution.  OpenFOAM's own field communication
handles the boundary-condition updates of \code{Pre} and \code{Pim} and the
evaluation of the derived fields.

\subsection{Installation and dependencies}
\label{sec:installation}

The complete fresh-clone procedure is documented in the repository file
\path{INSTALL_ACOUSTICS.md}; the essential steps are:
\begin{enumerate}[leftmargin=*]
    \item Install OpenFOAM-v2606 from openfoam.com and source its
    \code{etc/bashrc}.  Other OpenFOAM distributions are not supported.
    \item Clone the repository with \code{-{}-branch feature/acousticLevitation}
    (or check out revision \code{bb09dbfb}) and source \code{scripts/bashrc},
    which sets \code{TPF\_PROJECT\_DIR}, \code{PETSC\_DIR}, \code{PETSC\_ARCH}
    and the library path.
    \item Run \code{./petsc\_build.sh}.  The script clones PETSc at the pinned
    revision \code{0eed7b4d} (release 3.25.0) into \code{petsc/}, configures
    it with MPI, MUMPS 5.8.2, ScaLAPACK, METIS, ParMETIS, Hypre, SuperLU\_DIST
    and a reference BLAS/LAPACK, builds it in optimized mode with 32-bit
    indices, and optionally builds the OpenFOAM \code{petsc4Foam} module at
    revision \code{d72de946}.  The program links PETSc directly and does not
    require \code{petsc4Foam}: the supplied \code{controlDict} files list
    \code{libs ("libpetscFoam.so")}, but when that library is absent OpenFOAM
    only reports that it could not be loaded, and the computed fields are
    bit-identical.
    \item Run \code{./Allwmake}, which builds the \code{TwoPhaseFlow}
    libraries, the \code{acousticInterface} library, \code{setPMLFields}
    and the acoustic solvers, followed by
    \path{apps/utilities/preProcessing/setAlphaField/Allwmake}.
    \item Build the cfMesh plugin \cite{Juretic2015} shipped with OpenFOAM
    (\path{$WM_PROJECT_DIR/plugins/cfmesh/Allwmake}) if cases based on
    \code{cartesian2DMesh}, such as the levitator, are to be run.
    \item Run \code{scripts/check-acoustic-environment.sh}, which verifies the
    OpenFOAM version, the repository environment, the PETSc installation and
    the presence of all acoustic executables and their shared libraries
    without running a simulation.
\end{enumerate}
The verification-case post-processing requires Python~3 with NumPy, SciPy,
Matplotlib and VTK; \code{pytest} is needed for the test suite.  Compilers,
MPI (Open MPI \cite{Gabriel2004} in the tested configurations) and CMake are
provided by the operating system.  A smoke test with a coarse mesh,
\code{NX=400 ./Allrun} in the regression case directory, completes in seconds
and confirms that the build and the run-time environment are functional.

Two points observed during the independent rebuild of
\cref{sec:aiWorkflow} are relevant for installation.  First, the
\code{Make/options} file of the \code{acousticInterface} library in revision
\code{bb09dbfb} lists the user library \code{-lVoF} without adding
\path{-L$(FOAM_USER_LIBBIN)} to its link line, whereas the solver's
\code{Make/options} does.  On a system where the user library directory is
not otherwise on the linker search path, \code{./Allwmake} then fails to link
\code{libacousticInterface.so} and, consequently, \code{acousticHelmholtzFoam};
exporting \path{LIBRARY_PATH=$FOAM_USER_LIBBIN} before \code{./Allwmake}
resolves this, and adding the missing \code{-L} flag to that file removes the
dependence on the environment.  Second, the environment check requires the
cfMesh executables and the \code{petsc4Foam} checkout to be present and
reports their absence as failures, although neither is needed to build and
run the solver and the layered regression.  With OpenFOAM's default warning
flags, the acoustic targets compile with six old-style-cast warnings in the
PETSc interface code of \code{acousticHelmholtzFoam.C} and about 300 notes
that originate in PETSc headers; the \code{acousticInterface} library and
\code{setPMLFields} compile without warnings.

\subsection{Tests and continuous integration}
\label{sec:tests}

Three layers of tests are supplied.  The compiled unit checks in
\path{apps/benchmark/testAcousticInterface} exercise the geometric
face-cutting and transmission-fit routines against analytical configurations
and print \path{ACOUSTIC_INTERFACE_TESTS_PASSED} on success.  The
\code{pytest} module \path{testsuite/acousticInterface/test_invariants.py}
runs this binary, checks that invalid \code{acousticInterface}
configurations are rejected with the documented messages, verifies that a
missing sub-dictionary reproduces the explicit \code{legacy} setting to
$10^{-12}$, and confirms that the geometric face fraction is independent of
the sign of the face flux; it requires the compiled solver on the search path
and prepares a small layered case automatically.  The case-level \code{Allrun}
scripts are end-to-end regressions: the layered case of
\cref{sec:layeredRegression} compares against an analytical solution and
writes the error to \path{postProcessing/interfaceValidation/1/metrics.txt}.
Comparison drivers (\code{compare.py}, \code{advanced.py}, \code{wedge.py})
and audit scripts for the assembled operator and the shared processor
operators complete the suite; their results for the method paper are recorded
in \path{testsuite/acousticInterface/VERIFICATION.md}.

The repository's GitHub Actions workflow (\path{.github/workflows/openfoam.yml})
builds the framework and runs the \code{pytest} suite, but it is configured
for the \code{of2406} branch with a packaged OpenFOAM and without PETSc, so
it does not currently exercise the acoustic targets.  Automated coverage of
the acoustic solvers is therefore limited to the manual test layers above.

\section{Verification of the software workflow}
\label{sec:verification}

\subsection{Computational workflow and AI assistance}
\label{sec:aiWorkflow}

The convergence, PML-sensitivity, baffled-piston and radiation-force studies
that verify the numerical method are reported in \cite{Xu2026HelmholtzPML}
and are not repeated here.  This section retains one case as a reproducible
end-to-end software regression and adds the serial/parallel equivalence check
that is specific to the distributed assembly.  For this paper the program was
rebuilt from an independent clone of revision \code{bb09dbfb} with the
documented procedure of \cref{sec:installation} on a workstation with an
Intel Core Ultra~9 275HX processor (24 cores) running Ubuntu~24.04 under
WSL2, with GCC~13.3, Open~MPI~4.1.6, OpenFOAM-v2606 (build
\code{481094fdf3-20260618}) and PETSc~3.25.0 at the pinned revision; the
PETSc build took about four minutes with 20 parallel jobs and the OpenFOAM
targets a further four minutes.  All numbers in this section were obtained
with that build.  Claude (Anthropic) assisted the
corresponding author with the verification scripting, the comparison and
plotting code, and the consistency checks between manuscript, source code and
case files; the numerical results were produced by the compiled solver.

\subsection{Layered gas--liquid--PML regression}
\label{sec:layeredRegression}

The case \code{layeredInterface1D} combines a gas--liquid material jump, a
velocity-driven boundary and a directional PML in one dimension, and has an
analytical complex-pressure solution over the complete computational domain,
including the PML \cite{Xu2026HelmholtzPML}.  A transducer at $x=0$ with
velocity amplitude $u_0=0.01\,\mathrm{m\,s^{-1}}$ drives air
($\rho_1=1.2\,\mathrm{kg\,m^{-3}}$, $c_1=343\,\mathrm{m\,s^{-1}}$) at
$f=20\,\mathrm{kHz}$; the interface to water
($\rho_2=1000\,\mathrm{kg\,m^{-3}}$, $c_2=1500\,\mathrm{m\,s^{-1}}$) is at
$L_1=0.09\,\mathrm{m}$, the PML occupies $0.20\,\mathrm{m}<x<0.35\,\mathrm{m}$
with $\sigma_{\max}=5\times10^5\,\mathrm{s^{-1}}$ and $p_o=3$, and the outer
boundary is sound-hard.  The mesh has $N_x=8000$ uniform cells.  All
parameters are centralized in \code{caseParams.sh}; \code{prepareCase}
substitutes them into the mesh, material, boundary-condition and PML
dictionaries, and interchanging the two property pairs selects the reverse
water-to-air incidence reported in \cite{Xu2026HelmholtzPML}.

The \code{Allrun} script executes \code{blockMesh}, \code{setAlphaField}
(planar interface at $L_1$), \code{setPMLFields}, \code{decomposePar}, the
parallel solver, \code{reconstructPar} and the Python post-processing.  The
post-processing converts the cell fields to point data, probes them at 1200
uniformly spaced points over $0\le x\le L$, and evaluates the relative
complex-pressure error
$E_P=\|P_h-P\|_2/\|P\|_2$ with unit sample weights against the analytical
solution.  The reproduced value is $E_P=1.197\times10^{-3}$, identical to the
value reported in \cite{Xu2026HelmholtzPML} for this mesh.  The interface
$L_1$ lies inside a cell rather than on a face, so the regression exercises
the reconstructed face fraction of \cref{eq:faceInverseDensity}; this value is
therefore a workflow check and not the smallest attainable error.  Under
refinement with the interface and PML start aligned to mesh faces, $E_P$
decreases monotonically to $2.657\times10^{-4}$ at $N_x=8960$, and the
sensitivity to $\sigma_{\max}$ is tabulated in \cite{Xu2026HelmholtzPML}.
For the eight-rank run the PETSc diagnostics of \cref{sec:linearSolve}
report a relative residual of $5.5\times10^{-12}$ and a normwise backward
error of $5.9\times10^{-17}$; the relative residual is larger than the
backward error because the right-hand side is non-zero only in the driven
boundary cell, so that $\|\mathbf{b}\|_2\ll\|\mathbf{M}\|_\infty\|\mathbf{x}\|_\infty$.
The complete \code{Allrun}, including meshing, decomposition,
reconstruction and the Python post-processing, takes $2.8\,\mathrm{s}$ of
wall-clock time on the machine of \cref{sec:aiWorkflow}.

\subsection{Serial and MPI equivalence}
\label{sec:parallelEquivalence}

\begin{figure}[htbp]
    \centering
    \includegraphics[width=0.92\textwidth]{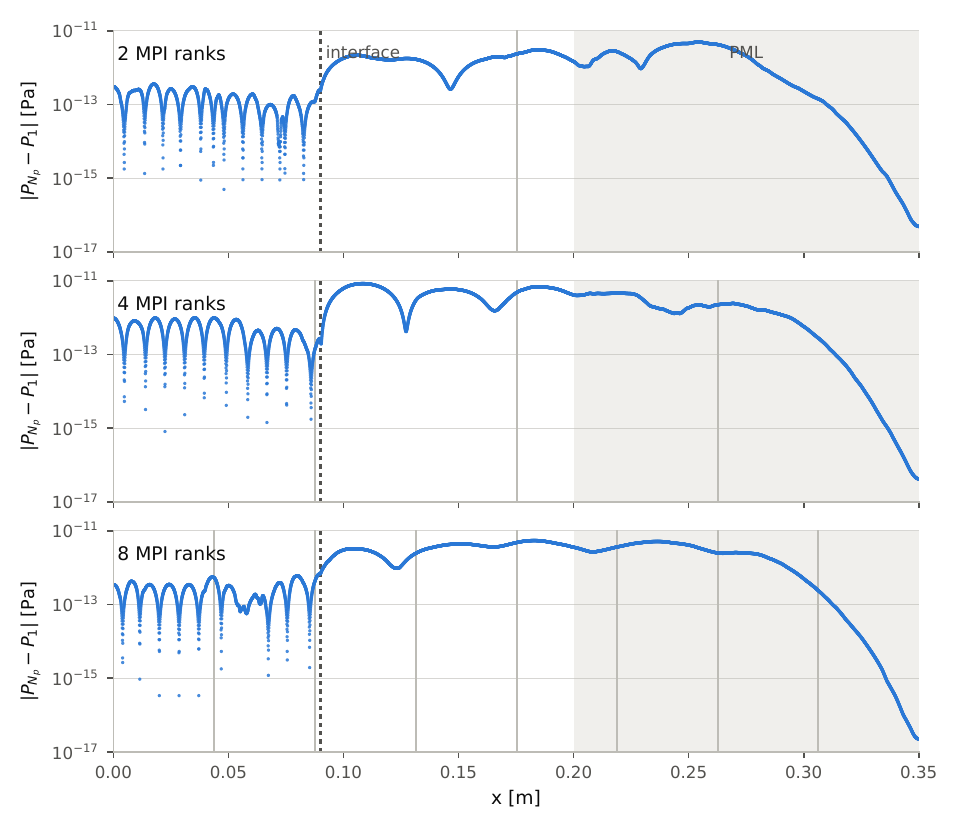}
    \caption{Per-cell modulus of the difference between the MPI and serial
    complex pressure for the layered regression case on 2, 4 and 8 ranks.
    Vertical grey lines mark the processor boundaries of each decomposition,
    the dotted line marks the gas--liquid interface and the shaded band is the
    PML.  \plotAidDisclosure}
    \label{fig:parallelAgreement}
\end{figure}

The distributed assembly of \cref{sec:assembly} is designed to produce the
same operator on any number of ranks.  To test this property, the regression
case was solved on one rank and on 2, 4 and 8 ranks with Scotch decomposition
and otherwise identical input, and the reconstructed cell fields were compared
in double precision; the fields were written in ASCII with 17 significant
digits for this comparison, because a comparison of fields written at the
case's default \code{writePrecision 10} is limited by output rounding to
about $10^{-10}$ relative and would hide the actual agreement.
\Cref{tab:serialParallelEquivalence} lists the maximum absolute and relative
$L_2$ differences, and \cref{fig:parallelAgreement} shows the per-cell
modulus of the complex difference along the domain together with the
processor boundaries.  The relative differences of $0.6$--$1.0\times10^{-12}$
are of the order of the relative residuals of the direct solves and originate
in the different summation orders of the distributed factorization; they show
no structure at the processor boundaries or at the material interface, which
confirms that the processor couplings, the boundary contributions and the
exchanged face fractions are assembled consistently.  The companion paper
reports the same property for the three-dimensional sphere and oblique-mesh
studies, where the geometric face fraction reduces the eight-rank difference
of the original donor-based face areas from $2.6\times10^{-6}$ to below
$10^{-14}$ \cite{Xu2026HelmholtzPML}.

\begin{table}[htbp]
    \centering
    \caption{Serial versus MPI equivalence for the layered regression case
    ($N_x=8000$, $\max|P|=4.12\,\mathrm{Pa}$).  Differences are evaluated
    cell by cell between the reconstructed parallel fields and the serial
    fields written with 17 significant digits; the relative $L_2$ difference
    is $\|\mathbf{P}_{N_p}-\mathbf{P}_1\|_2/\|\mathbf{P}_1\|_2$ for the complex
    pressure.  The last column is the normwise backward error of the PETSc
    solve reported by the solver (\cref{sec:linearSolve}).}
    \label{tab:serialParallelEquivalence}
    \small
    \setlength{\tabcolsep}{5pt}
    \begin{tabular}{rcccc}
        \toprule
        Ranks & Max.\ $|\Delta P_{re}|$ & Max.\ $|\Delta P_{im}|$ & Relative $L_2$ & Backward \\
        $N_p$ & [Pa] & [Pa] & difference & error \\
        \midrule
        1 & -- & -- & -- & $2.4\times10^{-16}$ \\
        2 & $3.99\times10^{-12}$ & $4.22\times10^{-12}$ & $5.66\times10^{-13}$ & $9.7\times10^{-17}$ \\
        4 & $5.95\times10^{-12}$ & $5.91\times10^{-12}$ & $1.03\times10^{-12}$ & $7.5\times10^{-18}$ \\
        8 & $4.60\times10^{-12}$ & $5.26\times10^{-12}$ & $8.47\times10^{-13}$ & $1.4\times10^{-16}$ \\
        \bottomrule
    \end{tabular}
\end{table}

\FloatBarrier
\section{Application example: levitator height sweep}
\label{sec:application}

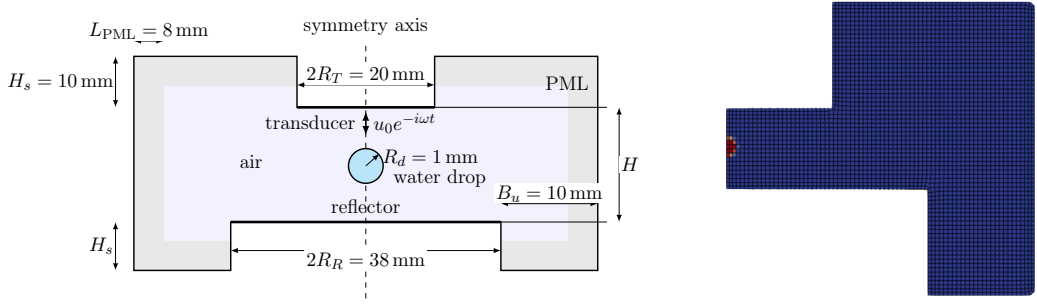
\begin{figure}[H]
    \centering
    \begin{minipage}[t]{0.62\textwidth}
    \centering
    \resizebox{\linewidth}{!}{%
    \begin{tikzpicture}[x=1cm,y=1cm,>=latex,font=\small]
        \def\xouter{4.55}
        \def\xref{2.65}
        \def\xtr{1.35}
        \def\pmlt{0.58}
        \def\ybot{-0.95}
        \def\yref{0.00}
        \def\ydrop{1.10}
        \def\yface{2.25}
        \def\ytop{3.25}
        \def\dropR{0.34}

        \fill[blue!5]
            (-\xouter,\ybot) -- (-\xouter,\ytop) -- (-\xtr,\ytop)
            -- (-\xtr,\yface) -- (\xtr,\yface) -- (\xtr,\ytop)
            -- (\xouter,\ytop) -- (\xouter,\ybot) -- (\xref,\ybot)
            -- (\xref,\yref) -- (-\xref,\yref) -- (-\xref,\ybot) -- cycle;

        \fill[gray!18] (-\xouter,\ybot) rectangle ({-\xouter+\pmlt},\ytop);
        \fill[gray!18] ({\xouter-\pmlt},\ybot) rectangle (\xouter,\ytop);
        \fill[gray!18] (-\xouter,{\ytop-\pmlt}) rectangle (-\xtr,\ytop);
        \fill[gray!18] (\xtr,{\ytop-\pmlt}) rectangle (\xouter,\ytop);
        \fill[gray!18] (-\xouter,\ybot) rectangle (-\xref,{\ybot+\pmlt});
        \fill[gray!18] (\xref,\ybot) rectangle (\xouter,{\ybot+\pmlt});

        \draw[thick]
            (-\xouter,\ybot) -- (-\xouter,\ytop) -- (-\xtr,\ytop)
            -- (-\xtr,\yface) -- (\xtr,\yface) -- (\xtr,\ytop)
            -- (\xouter,\ytop) -- (\xouter,\ybot) -- (\xref,\ybot)
            -- (\xref,\yref) -- (-\xref,\yref) -- (-\xref,\ybot) -- cycle;
        \draw[dashed] (0,{\ybot-0.55}) -- (0,{\ytop+0.28})
            node[above] {symmetry axis};
        \draw[line width=1.45pt] (-\xtr,\yface) -- (\xtr,\yface);
        \draw[<->,thick] (0,{\yface-0.08}) -- (0,{\yface-0.55})
            node[midway,right] {$u_0 e^{-i\omega t}$};
        \node[below left] at (-0.06,{\yface-0.02}) {transducer};
        \draw[line width=1.45pt] (-\xref,\yref) -- (\xref,\yref);
        \node[above] at (0,{\yref+0.03}) {reflector};
        \fill[cyan!25] (0,\ydrop) circle (\dropR);
        \draw[thick] (0,\ydrop) circle (\dropR);
        \draw[->] (0,\ydrop) -- (0.27,1.34)
            node[pos=0.68,right] {$R_d=1\,\mathrm{mm}$};
        \node[right] at (0.40,0.87) {water drop};
        \node at (-2.25,1.18) {air};
        \node at (3.95,2.72) {PML};

        \draw[<->] ({\xouter+0.42},\yref) -- ({\xouter+0.42},\yface)
            node[midway,right,fill=white,inner sep=1pt] {$H$};
        \draw (\xtr,\yface) -- ({\xouter+0.18},\yface);
        \draw (\xref,\yref) -- ({\xouter+0.18},\yref);
        \draw[<->] (-\xtr,{\yface+0.44}) -- (\xtr,{\yface+0.44})
            node[midway,above,fill=white,inner sep=1pt] {$2R_T=20\,\mathrm{mm}$};
        \draw[<->] (-\xref,{\yref-0.56}) -- (\xref,{\yref-0.56})
            node[midway,below,fill=white,inner sep=1pt] {$2R_R=38\,\mathrm{mm}$};
        \draw[<->] (\xref,0.36) -- (\xouter,0.36)
            node[midway,above,fill=white,inner sep=1pt] {$B_u=10\,\mathrm{mm}$};
        \draw[<->] (-\xouter,{\ytop+0.30}) -- ({-\xouter+\pmlt},{\ytop+0.30})
            node[midway,above,fill=white,inner sep=1pt] {$L_{\mathrm{PML}}=8\,\mathrm{mm}$};
        \draw[<->] ({-\xouter-0.36},\yface) -- ({-\xouter-0.36},\ytop)
            node[midway,left,fill=white,inner sep=1pt] {$H_s=10\,\mathrm{mm}$};
        \draw[<->] ({-\xouter-0.36},\ybot) -- ({-\xouter-0.36},\yref)
            node[midway,left,fill=white,inner sep=1pt] {$H_s$};
    \end{tikzpicture}
    }
    \end{minipage}
    \hfill
    \begin{minipage}[t]{0.31\textwidth}
    \centering
    \includegraphics[width=\linewidth]{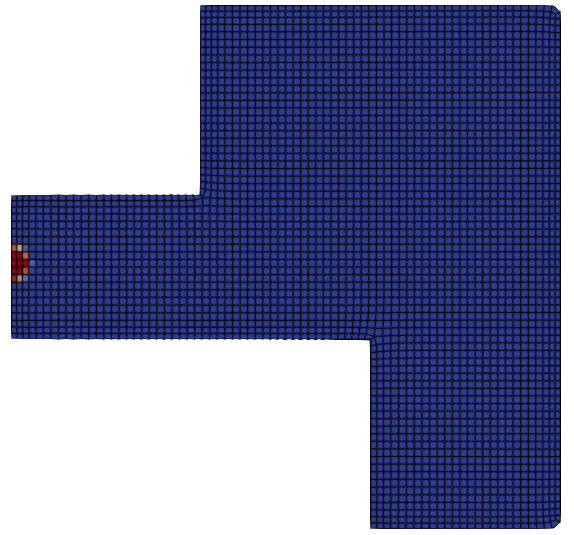}
    \end{minipage}
    \caption{Schematic meridional view of the single-axis levitator used in the
    application example.  The transducer--reflector spacing $H$ is swept, the
    reference drop is placed at mid-height, and the rectangular PML regions
    are indicated by the shaded bands.  The right panel shows the
    corresponding one-degree wedge mesh generated with cfMesh, with cells
    where $\alpha=1$ marked in red.}
    \label{fig:levitatorSketch}
\end{figure}

The application example reproduces the height-sweep configuration of
\cite{andrade2019numerical}: the radiation force on the reflector of a
single-axis levitator is evaluated while the transducer--reflector distance
$H$ is varied, for an empty levitator and for constant-volume spheroidal
drops of increasing aspect ratio.  \Cref{fig:levitatorSketch} shows the
geometry and the wedge mesh.

The case \code{WedgeLevitatorDropAlpha} is parameter-driven: all geometric,
material, PML, mesh and drop parameters are set in \code{caseParams.sh}, and
\code{prepareCase} renders the cfMesh surface, the mesh dictionaries,
\code{transportProperties} and the transducer boundary condition from them.
The \code{Allrun} script generates a two-dimensional cfMesh mesh
\cite{Juretic2015}, extrudes it into a one-degree wedge, initializes the drop
with \code{setAlphaField}, optionally refines the interface cells, runs
\code{setPMLFields} and solves in parallel.  The frequency is
$25.25\,\mathrm{kHz}$, the transducer velocity amplitude is
$1\,\mathrm{m\,s^{-1}}$, the air and water properties are
$\rho_g=1.2\,\mathrm{kg\,m^{-3}}$, $c_g=343\,\mathrm{m\,s^{-1}}$,
$\rho_l=998.3\,\mathrm{kg\,m^{-3}}$ and $c_l=1480\,\mathrm{m\,s^{-1}}$, and
the PML has $L=8\,\mathrm{mm}$, $\sigma_{\max}=2\times10^{6}\,\mathrm{s^{-1}}$
in the radial and axial directions, zero damping in the wedge direction
through \code{sigmaMaxVector}, and $p_o=3$.  The equivalent spherical drop
radius is $1\,\mathrm{mm}$; the drop is represented by \code{alpha.water}
with its volume held constant while the aspect ratio $a/b$ of the spheroid is
varied.  The script \code{runHeightSweep.sh} evaluates 100 uniformly spaced
values of $H$ between $0.46$ and $0.64$ wavelengths for the empty cavity and
for the aspect ratios $1$, $2$ and $3$, and writes the results to a CSV file
that \code{plotHeightSweep.py} plots.

The reflector force is evaluated by a coded function object in
\code{controlDict} that integrates the solver field \code{pr} over the
reflector patch,
\begin{equation}
    F_n = \mathbf{n}\cdot\sum_{f\in\Gamma_r} p_{\mathrm{rad},f}\,\mathbf{S}_f ,
    \label{eq:reflectorForce}
\end{equation}
where $\Gamma_r$ is the reflector patch and $\mathbf{n}$ its mean normal; the
momentum-flux contribution vanishes on the sound-hard reflector
(\cref{sec:derivedQuantities}).  The wedge force is multiplied by the
axisymmetric factor $360/\theta_{\mathrm{wedge}}=360$.  Since
\cref{eq:radiationPressure} is evaluated in the gas at the reflector, the
homogeneous-medium assumption of $p_{\mathrm{rad}}$ is satisfied where the
force is evaluated.

\begin{figure}[H]
    \centering
    \includegraphics[width=0.8\textwidth]{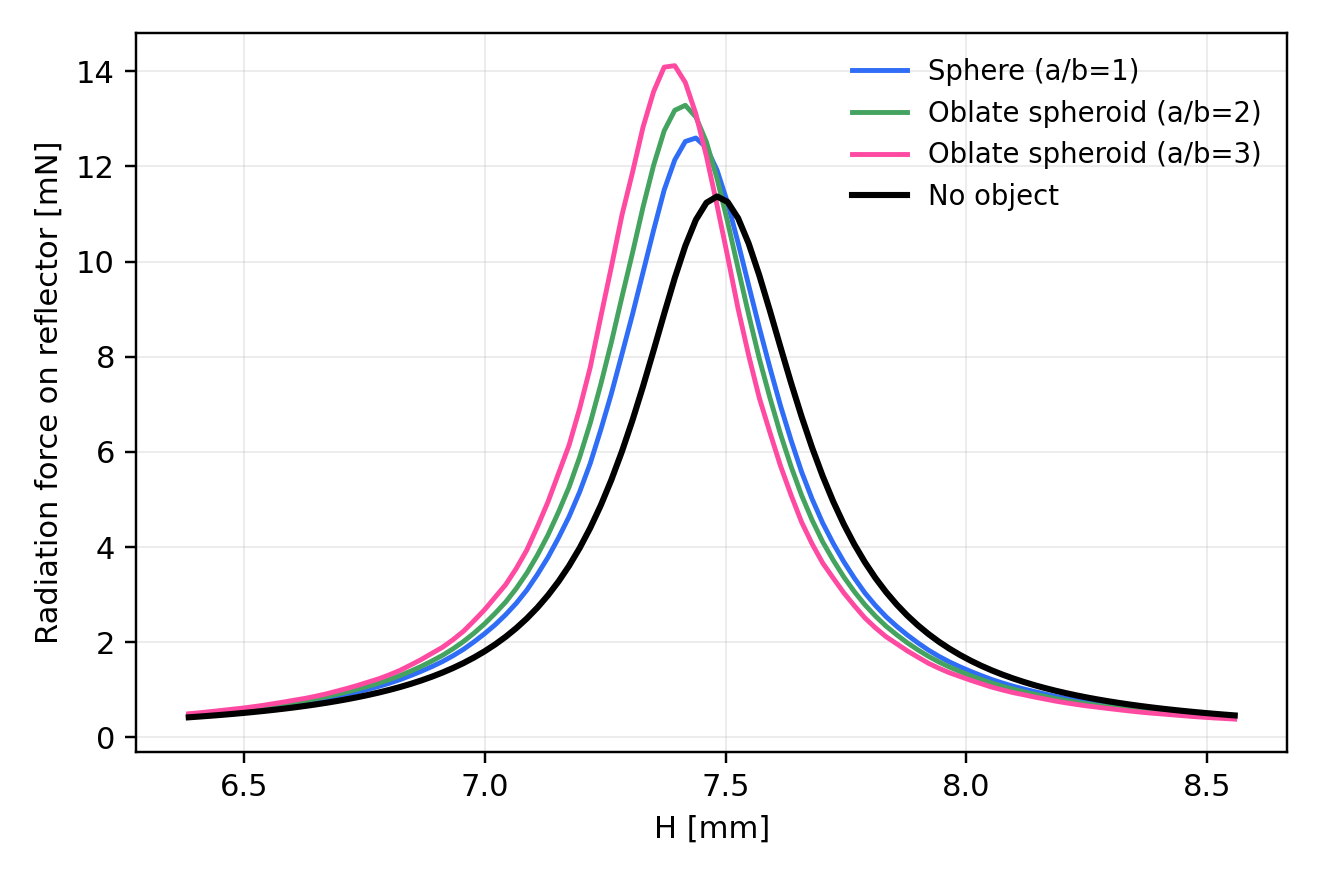}
    \caption{Application example: reflector radiation force during a height
    sweep of the wedge levitator for an empty cavity and constant-volume drops
    with different aspect ratios.  Each curve uses 100 height-sample points.}
    \label{fig:heightSweepAspectRatio}
\end{figure}

The curves in \cref{fig:heightSweepAspectRatio} reproduce the principal trends
of \cite{andrade2019numerical}: resonance amplification, a drop-induced shift
of the response, and increasing peak force with increasing aspect ratio.  The
example demonstrates a complete scripted sweep with a VOF-defined two-phase
material field, directional PML damping on a wedge mesh and solver-generated
acoustic-force output.  A quantitative comparison of the resolved forces with
the Gor'kov point-particle limit for rigid spheres is given in
\cite{Xu2026HelmholtzPML}.

\FloatBarrier
\section{Performance}
\label{sec:performance}

Performance was assessed with the \code{WedgeLevitatorDropAlpha} case, which
exercises the heterogeneous two-phase coefficients and the PML treatment of
the application example.  The measurements were collected on a workstation
with an AMD Ryzen~7 PRO 7840U processor (8 physical cores, 16 hardware
threads) with OpenFOAM-v2606.  The script \code{runPerformanceScaling.sh}
prepares each mesh once, performs one unreported warm-up solve so that the
compilation of the coded function object does not enter the timings, and
records the elapsed wall-clock time of each \code{acousticHelmholtzFoam}
invocation with \code{/usr/bin/time}, after mesh generation and, for MPI
runs, after domain decomposition.  Function objects were disabled during the
timed runs.  The solver assembles the block system with two unknowns per
cell and solves it with the default \code{preonly}/\code{lu}/MUMPS
configuration; the MPI runs used Scotch decomposition and Open MPI with
\code{-{}-oversubscribe}, as in the standard OpenFOAM run functions.

\begin{table}[H]
    \centering
    \caption{\code{WedgeLevitatorDropAlpha} strong-scaling timings for the
    $570\,818$-cell mesh, corresponding to $1\,141\,636$ block-system
    unknowns.}
    \label{tab:dropAlphaStrongScaling}
    \begin{tabular}{rrrr}
        \toprule
        MPI ranks & Elapsed time [s] & Speedup & Parallel efficiency [\%] \\
        \midrule
        1 & 16.77 & 1.00 & 100.0 \\
        2 & 12.13 & 1.38 & 69.1 \\
        4 & 10.25 & 1.64 & 40.9 \\
        8 & 10.96 & 1.53 & 19.1 \\
        \bottomrule
    \end{tabular}
\end{table}

\Cref{tab:dropAlphaStrongScaling} shows the strong-scaling behaviour for the
finest benchmark mesh, selected with \code{STRONG\_CELL\_SIZE=3e-5}.
Increasing the rank count from one to eight reduces the elapsed time from
$16.77\,\mathrm{s}$ to $10.96\,\mathrm{s}$; the minimum of $10.25\,\mathrm{s}$
is reached at four ranks, and the eight-rank efficiency falls to $19.1\%$.
This behaviour is expected for a parallel sparse direct factorization of a
workstation-scale problem: the factorization of the two-dimensional wedge
system is dominated by dense frontal-matrix operations whose parallel
efficiency on a single memory-bandwidth-limited processor saturates quickly,
and the symbolic analysis and ordering phases are largely sequential.  The
timings characterize the default direct-solver workflow on a workstation; they
are not a scalability claim for large core counts.

\begin{figure}[H]
    \centering
    \includegraphics[width=0.92\textwidth]{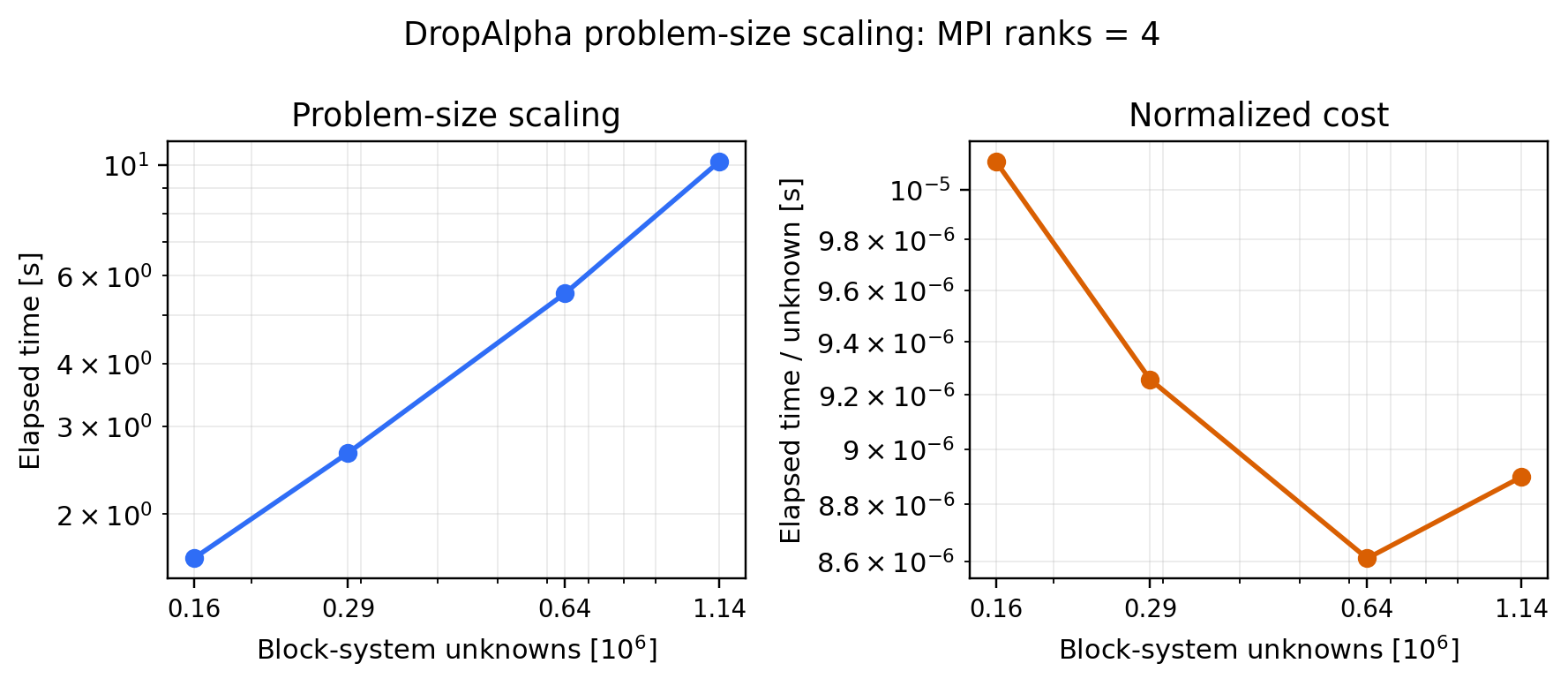}
    \caption{\code{WedgeLevitatorDropAlpha} problem-size scaling at four MPI
    ranks.  The mesh size is controlled by the cfMesh maximum cell size, and
    the plotted unknown count is twice the cell count.}
    \label{fig:dropAlphaProblemSizeScaling}
\end{figure}

\cref{fig:dropAlphaProblemSizeScaling} reports the fixed-rank problem-size
study.  With four MPI ranks, decreasing the cfMesh maximum cell size from
$8.0\times10^{-5}\,\mathrm{m}$ to $3.0\times10^{-5}\,\mathrm{m}$ increases the
mesh from $80\,569$ to $570\,818$ cells and the block system from $161\,138$ to
$1\,141\,636$ unknowns.  Over this range the measured elapsed time increases
from $1.63\,\mathrm{s}$ to $10.16\,\mathrm{s}$.  Thus the unknown count
increases by a factor of $7.08$ while the elapsed time increases by a factor of
$6.23$; a log-log fit gives an empirical exponent of approximately $0.93$.
The cost per unknown remains between $8.61\times10^{-6}$ and
$1.01\times10^{-5}\,\mathrm{s}$.  For these moderately sized axisymmetric
wedge meshes the direct factorization therefore grows close to linearly with
problem size, which is consistent with the nearly one-dimensional connectivity
of a thin wedge mesh.  The complexity of a sparse direct solve is superlinear
for genuinely three-dimensional meshes, and memory rather than time is then
the limiting resource; iterative Krylov methods with the preconditioners
available in the PETSc build can be selected through PETSc options
(\cref{sec:linearSolve}), but standard algebraic preconditioners are not
expected to converge for the indefinite Helmholtz--PML system without
Helmholtz-specific preconditioning, as the BoomerAMG test of
\cref{sec:linearSolve} illustrates, and no systematic assessment has been
made with this software.

\section{Availability, reproducibility and restrictions}
\label{sec:availability}

The program is distributed under the GNU General Public License version~3
(file \code{LICENSE} of the repository) in the \code{TwoPhaseFlow}
repository, \url{https://github.com/tmaric/TwoPhaseFlow}, on the branch
\code{feature/acousticLevitation}.  The revision described in this paper and
used for the companion paper is
\longhash{bb09dbfb34c1830478feecd8dd445957ac15949c}
\cite{XuLiuMaric2026AcousticSoftware}; the subsequent commits on the branch
up to \code{4bc88bdb} add the secondary-data package and provenance records
of the companion paper (\path{data/publications/helmholtz-pml/},
\code{PROVENANCE.md}) and leave the program sources unchanged.  The
secondary data of the companion paper are archived in TUdatalib
\cite{XuMaric2026HelmholtzPMLData}.  The program package for the CPC Program
Library contains the solver and utility sources, the \code{acousticInterface}
library, the \code{setAlphaField} utility, the build and environment scripts,
the two cases of this paper with their \code{Allrun} scripts, and the
reference output of the layered regression.

The regression is reproduced with \code{./Allrun} in
\path{run/acousticTests/FrequencyDomainTests/layeredInterface1D}; the
expected error appears as \code{P\_relL2} in
\path{postProcessing/interfaceValidation/1/metrics.txt}.  The height sweep
is reproduced with \code{RUN\_MODE=fullFig4Sweep ./Allrun} and the timing
study with \code{./runPerformanceScaling.sh} in the levitator case directory;
the strong-scaling mesh is selected with \code{STRONG\_CELL\_SIZE=3e-5} and
the problem-size study uses the four cell sizes of
\cref{fig:dropAlphaProblemSizeScaling} at four ranks.  Absolute timings
depend on the hardware.

The following restrictions apply to the present revision:
\begin{itemize}[leftmargin=*]
    \item Linear, inviscid, adiabatic acoustics about a quiescent state; no
    mean flow, viscosity, heat conduction or nonlinear effects.  The
    interface and the material properties are fixed during a solve.
    \item Separable Cartesian rectangular PMLs only; material interfaces must
    lie in the physical domain and each PML must contain a homogeneous
    continuation of the adjacent medium.
    \item The radiation pressure of \cref{eq:radiationPressure} is exact only
    where the medium is locally homogeneous.
    \item The default direct solver limits the problem size by memory; each
    non-orthogonal correction sweep refactorizes the matrix.
    \item Moving meshes and cyclic, AMI and overset patches are rejected by
    the geometric interface treatment; the experimental
    \code{plicTransmission} flux is not validated.
    \item Tested only with OpenFOAM-v2606 from openfoam.com.
\end{itemize}

\section{Conclusions}
\label{sec:conclusions}

\code{acousticHelmholtzFoam} provides frequency-domain acoustics for
heterogeneous gas--liquid media inside the OpenFOAM ecosystem.  It combines
volume-fraction-based material fields with a geometric face-fraction closure,
separable rectangular PML generation by \code{setPMLFields}, explicit assembly
of the real--imaginary block system into a distributed PETSc matrix, and a
MUMPS direct solve, and it writes pressure, velocity, radiation-pressure and
momentum-flux fields for subsequent force evaluation.  This paper documented
the implemented equations and boundary conditions, the complete case
interface including the \code{acousticInterface} selection, the assembly and
solution algorithm, the build procedure and the tests.

Rebuilt from an independent clone, the layered regression reproduces the
whole-domain complex-pressure error of $1.197\times10^{-3}$ on 8000 cells, and
solutions on one, two, four and eight ranks agree to within $10^{-12}$ in
relative $L_2$ norm, which confirms the consistency of the distributed
assembly.  The levitator height sweep
demonstrates parameter-driven case generation with a VOF-defined drop and
reproduces the benchmark trends in resonance position and force ordering.  For
the $570\,818$-cell levitator mesh the direct solve takes between
$16.77\,\mathrm{s}$ on one rank and $10.25\,\mathrm{s}$ on four ranks, and at
fixed rank count the time grows approximately linearly with the block-system
size over the tested range.  Future work will assess preconditioned iterative
solvers for larger three-dimensional problems, extend the automated tests to
the acoustic targets, and couple the solver more tightly to interface-resolved
two-phase flow simulations.

\section*{Code and data availability}

The program sources are available at
\url{https://github.com/tmaric/TwoPhaseFlow} on the branch
\code{feature/acousticLevitation}; the revision described in this paper is
\longhash{bb09dbfb34c1830478feecd8dd445957ac15949c}
\cite{XuLiuMaric2026AcousticSoftware}, and the program package will be
archived in the CPC Program Library.  The cases of \cref{sec:verification} and
\cref{sec:application} are included in the repository under
\path{run/acousticTests/FrequencyDomainTests/}.  The secondary data of the
companion method paper, including the layered-case profiles, are deposited in
TUdatalib \cite{XuMaric2026HelmholtzPMLData} (DOI registration pending at the
time of writing).

\section*{Acknowledgements}

The authors gratefully acknowledge the computing time provided on the
high-performance computer Lichtenberg~II at TU Darmstadt, funded by the German
Federal Ministry of Research, Technology and Space (BMFTR), and the State of
Hesse Ministry of Science and Research, Art and Culture (HMWK).

\section*{Funding}

Dieter Bothe acknowledges support through the tandem project with Mario Kupnik
funded by the Deutsche Forschungsgemeinschaft (DFG, German Research
Foundation) -- project number 542327521.  The funder had no role in software
design, data collection, analysis or interpretation, manuscript preparation,
or the decision to submit the article for publication.

\section*{Declaration of competing interest}

The authors declare that they have no known competing financial interests or
personal relationships that could have appeared to influence the work reported
in this paper.

\section*{CRediT authorship contribution statement}

\noindent\textbf{Chuanchao Xu}: Conceptualization, Software, Validation,
Investigation, Data curation, Visualization, Writing -- original draft.

\noindent\textbf{Jun Liu}: Software, Methodology, Investigation, Writing --
review \& editing.

\noindent\textbf{Dieter Bothe}: Conceptualization, Methodology, Resources,
Supervision, Project administration, Writing -- review \& editing.

\noindent\textbf{Tomislav Maric}: Conceptualization, Software, Methodology,
Validation, Supervision, Writing -- review \& editing.

\section*{Declaration of generative AI and AI-assisted technologies in the manuscript preparation process}

During the preparation of this work the authors used Claude (Anthropic) under
the direction of the corresponding author to assist with the revision of the
manuscript, with consistency checks between the manuscript, the source code
and the case files, and with the scripts of the independent rebuild and
serial/parallel comparison of \cref{sec:verification}.  The development of the
software and of the verification cases in the repository was assisted by
OpenAI Codex, as recorded in the repository's provenance documents and in the
companion paper.  After using these tools, the authors reviewed and edited the
content as needed and take full responsibility for the content of the
publication.

\bibliographystyle{elsarticle-num}
\bibliography{acousticHelmholtzFoam-revised}

\begin{thebibliography}{10}
\expandafter\ifx\csname url\endcsname\relax
  \def\url#1{\texttt{#1}}\fi
\expandafter\ifx\csname urlprefix\endcsname\relax\def\urlprefix{URL }\fi
\expandafter\ifx\csname href\endcsname\relax
  \def\href#1#2{#2} \def\path#1{#1}\fi

\bibitem{Andrade2018review}
M.~A. Andrade, N.~P{\'e}rez, J.~C. Adamowski,
  \href{https://link.springer.com/article/10.1007/s13538-017-0552-6}{Review of
  progress in acoustic levitation}, Brazilian Journal of Physics 48 (2018)
  190--213.
\newline\urlprefix\url{https://link.springer.com/article/10.1007/s13538-017-0552-6}

\bibitem{andrade2019numerical}
M.~A. Andrade, A.~Marzo, Numerical and experimental investigation of the
  stability of a drop in a single-axis acoustic levitator, Physics of Fluids
  31~(11) (2019) 117101.
\newblock \href {https://doi.org/10.1063/1.5121728}
  {\path{doi:10.1063/1.5121728}}.

\bibitem{Weller1998}
H.~G. Weller, G.~Tabor, H.~Jasak, C.~Fureby, A tensorial approach to
  computational continuum mechanics using object-oriented techniques, Computers
  in Physics 12~(6) (1998) 620--631.
\newblock \href {https://doi.org/10.1063/1.168744}
  {\path{doi:10.1063/1.168744}}.

\bibitem{Balay1997}
S.~Balay, W.~D. Gropp, L.~C. McInnes, B.~F. Smith, Efficient management of
  parallelism in object-oriented numerical software libraries, in: E.~Arge,
  A.~M. Bruaset, H.~P. Langtangen (Eds.), Modern Software Tools in Scientific
  Computing, Birkh{\"a}user, Boston, 1997, pp. 163--202.
\newblock \href {https://doi.org/10.1007/978-1-4612-1986-6_8}
  {\path{doi:10.1007/978-1-4612-1986-6_8}}.

\bibitem{PETScManual}
S.~Balay, S.~Abhyankar, M.~F. Adams, S.~Benson, J.~Brown, P.~Brune,
  K.~Buschelman, E.~M. Constantinescu, L.~Dalcin, A.~Dener, V.~Eijkhout,
  J.~Faibussowitsch, W.~D. Gropp, V.~Hapla, T.~Isaac, P.~Jolivet, D.~Karpeev,
  D.~Kaushik, M.~G. Knepley, F.~Kong, S.~Kruger, D.~A. May, L.~C. McInnes,
  R.~T. Mills, L.~Mitchell, T.~Munson, J.~E. Roman, K.~Rupp, P.~Sanan,
  J.~Sarich, B.~F. Smith, H.~Suh, S.~Zampini, H.~Zhang, J.~Zhang,
  \href{https://petsc.org/release/docs/manual/}{{PETSc/TAO} users manual},
  Argonne National Laboratory, Technical Report ANL-21/39, Revision 3.25
  (2026).
\newline\urlprefix\url{https://petsc.org/release/docs/manual/}

\bibitem{Amestoy2001}
P.~R. Amestoy, I.~S. Duff, J.-Y. L'Excellent, J.~Koster, A fully asynchronous
  multifrontal solver using distributed dynamic scheduling, SIAM Journal on
  Matrix Analysis and Applications 23~(1) (2001) 15--41.
\newblock \href {https://doi.org/10.1137/S0895479899358194}
  {\path{doi:10.1137/S0895479899358194}}.

\bibitem{Amestoy2019}
P.~R. Amestoy, A.~Buttari, J.-Y. L'Excellent, T.~Mary, Performance and
  scalability of the block low-rank multifrontal factorization on multicore
  architectures, ACM Transactions on Mathematical Software 45~(1) (2019)
  2:1--2:26.
\newblock \href {https://doi.org/10.1145/3242094} {\path{doi:10.1145/3242094}}.

\bibitem{ScheuflerRoenby2023TwoPhaseFlow}
H.~Scheufler, J.~Roenby, {TwoPhaseFlow}: A framework for developing two phase
  flow solvers in {OpenFOAM}, OpenFOAM Journal 3 (2023) 200--224.
\newblock \href {https://doi.org/10.51560/ofj.v3.80}
  {\path{doi:10.51560/ofj.v3.80}}.

\bibitem{Roenby2016}
J.~Roenby, H.~Bredmose, H.~Jasak, A computational method for sharp interface
  advection, Royal Society Open Science 3~(11) (2016) 160405.
\newblock \href {https://doi.org/10.1098/rsos.160405}
  {\path{doi:10.1098/rsos.160405}}.

\bibitem{Scheufler2019}
H.~Scheufler, J.~Roenby, Accurate and efficient surface reconstruction from
  volume fraction data on general meshes, Journal of Computational Physics 383
  (2019) 1--23.
\newblock \href {https://doi.org/10.1016/j.jcp.2019.01.009}
  {\path{doi:10.1016/j.jcp.2019.01.009}}.

\bibitem{Nicoud2007}
F.~Nicoud, L.~Benoit, C.~Sensiau, T.~Poinsot, Acoustic modes in combustors with
  complex impedances and multidimensional active flames, AIAA Journal 45~(2)
  (2007) 426--441.
\newblock \href {https://doi.org/10.2514/1.24933} {\path{doi:10.2514/1.24933}}.

\bibitem{Long2015}
W.~Long, Z.~Yang, A.~E. Copping, K.~W. Jung, Z.~Deng, The development of a
  finite volume method for modeling sound in coastal ocean environment, in:
  OCEANS 2015 -- MTS/IEEE Washington, 2015, pp. 1--6.
\newblock \href {https://doi.org/10.23919/OCEANS.2015.7404439}
  {\path{doi:10.23919/OCEANS.2015.7404439}}.

\bibitem{Harari2000}
I.~Harari, M.~Slavutin, E.~Turkel, Analytical and numerical studies of a finite
  element {PML} for the {Helmholtz} equation, Journal of Computational
  Acoustics 8~(1) (2000) 121--137.
\newblock \href {https://doi.org/10.1142/S0218396X0000008X}
  {\path{doi:10.1142/S0218396X0000008X}}.

\bibitem{Bermudez2007}
A.~Berm{\'u}dez, L.~Hervella-Nieto, A.~Prieto, R.~Rodr{\'i}guez, An optimal
  perfectly matched layer with unbounded absorbing function for time-harmonic
  acoustic scattering problems, Journal of Computational Physics 223~(2) (2007)
  469--488.
\newblock \href {https://doi.org/10.1016/j.jcp.2006.09.018}
  {\path{doi:10.1016/j.jcp.2006.09.018}}.

\bibitem{Modave2014PML}
A.~Modave, E.~Delhez, C.~Geuzaine, Optimizing perfectly matched layers in
  discrete contexts, International Journal for Numerical Methods in Engineering
  99~(6) (2014) 410--437.
\newblock \href {https://doi.org/10.1002/nme.4690}
  {\path{doi:10.1002/nme.4690}}.

\bibitem{Xu2026HelmholtzPML}
C.~Xu, J.~Liu, J.~H. D{\"o}rsam, M.~Kupnik, T.~Maric, D.~Bothe, An unstructured
  finite-volume {Helmholtz} method with perfectly matched layers for
  heterogeneous two-phase acoustics, arXiv:2610.01593 [physics.flu-dyn] (2026).
\newblock \href {https://doi.org/10.48550/arXiv.2610.01593}
  {\path{doi:10.48550/arXiv.2610.01593}}.

\bibitem{Jasak1996}
H.~Jasak, \href{http://hdl.handle.net/10044/1/8335}{Error analysis and
  estimation for the finite volume method with applications to fluid flows},
  Ph.D. thesis, Imperial College, University of London (1996).
\newline\urlprefix\url{http://hdl.handle.net/10044/1/8335}

\bibitem{Moukalled2016}
F.~Moukalled, L.~Mangani, M.~Darwish, The Finite Volume Method in Computational
  Fluid Dynamics: An Advanced Introduction with {OpenFOAM} and {MATLAB}, Vol.
  113 of Fluid Mechanics and Its Applications, Springer, Cham, 2016.

\bibitem{bruus2012acoustofluidicsARF}
H.~Bruus, Acoustofluidics 7: The acoustic radiation force on small particles,
  Lab on a Chip 12~(6) (2012) 1014--1021.

\bibitem{Chevalier2008}
C.~Chevalier, F.~Pellegrini, {PT-Scotch}: A tool for efficient parallel graph
  ordering, Parallel Computing 34~(6--8) (2008) 318--331.
\newblock \href {https://doi.org/10.1016/j.parco.2007.12.001}
  {\path{doi:10.1016/j.parco.2007.12.001}}.

\bibitem{Juretic2015}
F.~Jureti{\'c}, {cfMesh} User Guide, Creative Fields Ltd., Zagreb, version 1.1
  (2015).

\bibitem{Gabriel2004}
E.~Gabriel, G.~E. Fagg, G.~Bosilca, T.~Angskun, J.~J. Dongarra, J.~M. Squyres,
  V.~Sahay, P.~Kambadur, B.~Barrett, A.~Lumsdaine, R.~H. Castain, D.~J. Daniel,
  R.~L. Graham, T.~S. Woodall, {Open MPI}: Goals, concept, and design of a next
  generation {MPI} implementation, in: Recent Advances in Parallel Virtual
  Machine and Message Passing Interface (EuroPVM/MPI 2004), Vol. 3241 of
  Lecture Notes in Computer Science, Springer, 2004, pp. 97--104.
\newblock \href {https://doi.org/10.1007/978-3-540-30218-6_19}
  {\path{doi:10.1007/978-3-540-30218-6_19}}.

\bibitem{XuLiuMaric2026AcousticSoftware}
C.~Xu, J.~Liu, T.~Maric,
  \href{https://github.com/tmaric/TwoPhaseFlow/tree/bb09dbfb34c1830478feecd8dd445957ac15949c}{{acousticHelmholtzFoam}:
  geometric face averaging with the retained acoustic flux}, [software]
  {TwoPhaseFlow}, {GitHub}, revision {bb09dbfb34c1}, 9 September 2026 (2026).
\newline\urlprefix\url{https://github.com/tmaric/TwoPhaseFlow/tree/bb09dbfb34c1830478feecd8dd445957ac15949c}

\bibitem{XuMaric2026HelmholtzPMLData}
C.~Xu, T.~Maric,
  \href{https://tudatalib.ulb.tu-darmstadt.de/handle/tudatalib/5583}{An
  unstructured finite-volume {Helmholtz} method with perfectly matched layers
  for heterogeneous two-phase acoustics - secondary data
  ({acousticHelmholtzFoam} revision {bb09dbfb34c1}, dataset v1.0)}, [dataset]
  {TUdatalib}, version 1.0, {DOI} reserved; registration pending (2026).
\newblock \href {https://doi.org/10.48328/tudatalib-2357}
  {\path{doi:10.48328/tudatalib-2357}}.
\newline\urlprefix\url{https://tudatalib.ulb.tu-darmstadt.de/handle/tudatalib/5583}

\end{thebibliography}

\end{document}